\documentclass[aps,prl,twocolumn,superscriptaddress,calc,epsfig,10pt]{revtex4-1}
\usepackage{graphicx}
\usepackage{graphics}
\usepackage{amssymb,amsmath}
\usepackage{comment}

\usepackage{wasysym}

\usepackage{color}
\usepackage[normalem]{ulem}

\newcommand{\CUAaddress}{Harvard-MIT Center for Ultracold Atoms, Cambridge, Massachusetts 02138, USA}
\newcommand{\HarvardPhysicsAddress}{Department of Physics, Harvard University, Cambridge, Massachusetts 02138, USA}
\newcommand{\HarvardChemistryaddress}{Department of Chemistry and Chemical Biology, Harvard University, Cambridge, Massachusetts 02138, USA}
\newcommand{\NISTaddress}{Current address: Time and Frequency Division, National
Institute of Standards and Technology, Boulder, Colorado 80305, USA}
\newcommand{\QuEraaddress}{Current address: QuEra Computing, 1284 Soldiers Field Road, Boston, Massachusetts 02135, USA}

\newcommand{\kt}[1]{\ensuremath{\left|#1\right\rangle}}

\begin{document}

\title{Detection of Long-Lived Complexes in Ultracold Atom-Molecule Collisions}

\author{Matthew A. Nichols}
\email{matthewnichols@g.harvard.edu}
\affiliation{\HarvardChemistryaddress}
\affiliation{\HarvardPhysicsAddress}
\affiliation{\CUAaddress}

\author{Yi-Xiang Liu}
\affiliation{\HarvardChemistryaddress}
\affiliation{\HarvardPhysicsAddress}
\affiliation{\CUAaddress}

\author{Lingbang Zhu}
\affiliation{\HarvardChemistryaddress}
\affiliation{\HarvardPhysicsAddress}
\affiliation{\CUAaddress}

\author{Ming-Guang Hu}
\altaffiliation{\QuEraaddress}
\affiliation{\HarvardChemistryaddress}
\affiliation{\HarvardPhysicsAddress}
\affiliation{\CUAaddress}

\author{Yu Liu}
\altaffiliation{\NISTaddress}
\affiliation{\HarvardChemistryaddress}
\affiliation{\HarvardPhysicsAddress}
\affiliation{\CUAaddress}

\author{Kang-Kuen Ni}
\email{ni@chemistry.harvard.edu}
\affiliation{\HarvardChemistryaddress}
\affiliation{\HarvardPhysicsAddress}
\affiliation{\CUAaddress}
\date{\today}

\begin{abstract}
We investigate collisional loss in an ultracold mixture of $^{40}$K$^{87}$Rb molecules and $^{87}$Rb atoms, where chemical reactions between the two species are energetically forbidden. Through direct detection of the KRb$_{2}^{*}$ intermediate complexes formed from atom-molecule collisions, we show that a $1064$ nm laser source used for optical trapping of the sample can efficiently deplete the complex population via photo-excitation, an effect which can explain the universal two-body loss observed in the mixture. By monitoring the time-evolution of the KRb$_{2}^{*}$ population after a sudden reduction in the $1064$ nm laser intensity, we measure the lifetime of the complex ($0.39(6)$ ms), as well as the photo-excitation rate for $1064$ nm light ($0.50(3)$ $\mu$s$^{-1}($kW/cm$^{2})^{-1}$). The observed lifetime is ${\sim}10^5$ times longer than recent estimates based on the Rice–Ramsperger–Kassel–Marcus statistical theory, which calls for new insight to explain such a dramatic discrepancy.
\end{abstract}

\maketitle

At the microscopic level, chemical reactions are quantum mechanical transformations from one chemical species to another. Establishing a full quantum picture of such processes therefore requires the ability to resolve the relevant quantum states of these species. With the development of techniques to create cold and ultracold molecular samples, it has become possible to control all quantum degrees of freedom of molecular reactants~\cite{quemener2012ultracold,Jankunas2015,Bohn2017Review}, which has allowed for precise investigations of molecular collisions and chemistry in the lowest temperature regimes~\cite{yang2007state,balakrishnan2016perspective,Dulieu2018Book,toscano2020cold, liu2020probing}. Previous studies, for instance, have elucidated the various roles which scattering resonances~\cite{klein2017directly,yang2019observation,de2020imaging} and long-range interactions~\cite{ospelkaus2010quantum,ni2010dipolar,hall2012millikelvin,perreault2017quantum,guo2018dipolar,kilaj2018observation,puri2019reaction} play in determining reaction rates, and have even demonstrated the ability to control the quantum states of reaction outcomes~\cite{hu2020product}.

In the particular instance of ultracold collisions between bialkali dimers, one prominent feature of the corresponding short-range collision dynamics is the formation of long-lived intermediate complexes~\cite{hu2019direct}. These complexes can live for millions of molecular vibrations due to the limited number of dissociation channels that are available when the reactants are prepared in their lowest energy quantum states~\cite{mayle2012statistical,mayle2013scattering}. Because of this, they not only impact the quantum state distribution of reaction products, as they can redistribute energy among the various modes of motion~\cite{bonnet1999some}, but they can also dramatically affect the stability of molecular gases where exothermic reaction channels do not exist. This can happen, for instance, if the complexes are excited by photons from external laser sources~\cite{christianen2019photoinduced,liu2020photo,gregory2020loss}, or if they undergo collisions with other atoms or molecules~\cite{mayle2012statistical,mayle2013scattering,croft2014long}. Due to their extremely long lifetimes, however, a full theoretical description of the resulting highly convoluted collision dynamics is immensely challenging~\cite{croft2014long,li2020advances}. Nevertheless, intimate details of such dynamics can be observed experimentally. Recently, an experimental study of the product quantum state distribution for the ultracold reaction, $^{40}$K$^{87}$Rb$\,+\,^{40}$K$^{87}$Rb$\,\rightarrow\,$K$_{2}$Rb$_{2}^{*}\rightarrow\,$K$_{2}\,+\,$Rb$_{2}$, rigorously probed for statistical behavior~\cite{light1967statistical,pechukas1976statistical,nikitin2012theory}, and revealed deviations whose explanation requires computational capabilities beyond the current state-of-the-art~\cite{liu2020precision}. To gain further insight into the properties of these systems, it is  desirable to find related situations where quantum dynamics calculations are feasible, so that one can benchmark theoretical predictions using experimental measurements. One way to achieve this is to investigate collisions involving fewer atoms, such as those which occur between ultracold bialkali dimers and  alkali atoms. In this case, the limited atom number within the associated collision complex reduces the computational complexity, so that calculations of the corresponding collision dynamics are potentially within reach of current theoretical techniques and computational powers~\cite{croft2017universality,Kendrick2021}.

In this work, we experimentally examine the collisional properties of the KRb-Rb system. Despite the fact that the two-body chemical reaction is energetically forbidden, we observe universal two-body loss of the atoms and molecules. To investigate its origin, we directly probe the collision complexes, KRb$_{2}^{*}$, using ionization detection. We observe a severe reduction in the complex population with increasing intensities of our $1064$ nm trapping light. By rapidly changing the $1064$ nm intensity, we actively induce dynamics in the population to measure the complex lifetime and the first-order photo-excitation rate constant, which are $0.39(6)$ ms and $0.50(3)$ $\mu$s$^{-1}($kW/cm$^{2})^{-1}$, respectively. This lifetime is approximately five orders of magnitude larger than the most recent theoretical predictions~\cite{PrivateCommunication,ChristianenDOS2019}, which are based on the Rice–Ramsperger–Kassel–Marcus (RRKM) statistical theory~\cite{levine2009molecular}. Such an enormous discrepancy demands a critical examination of the current theoretical framework for ultracold molecular scattering, even for the case of collisions involving only three atoms, to find what might be missing.

\begin{figure}[t]
\centering
\includegraphics[width = 0.45\textwidth]{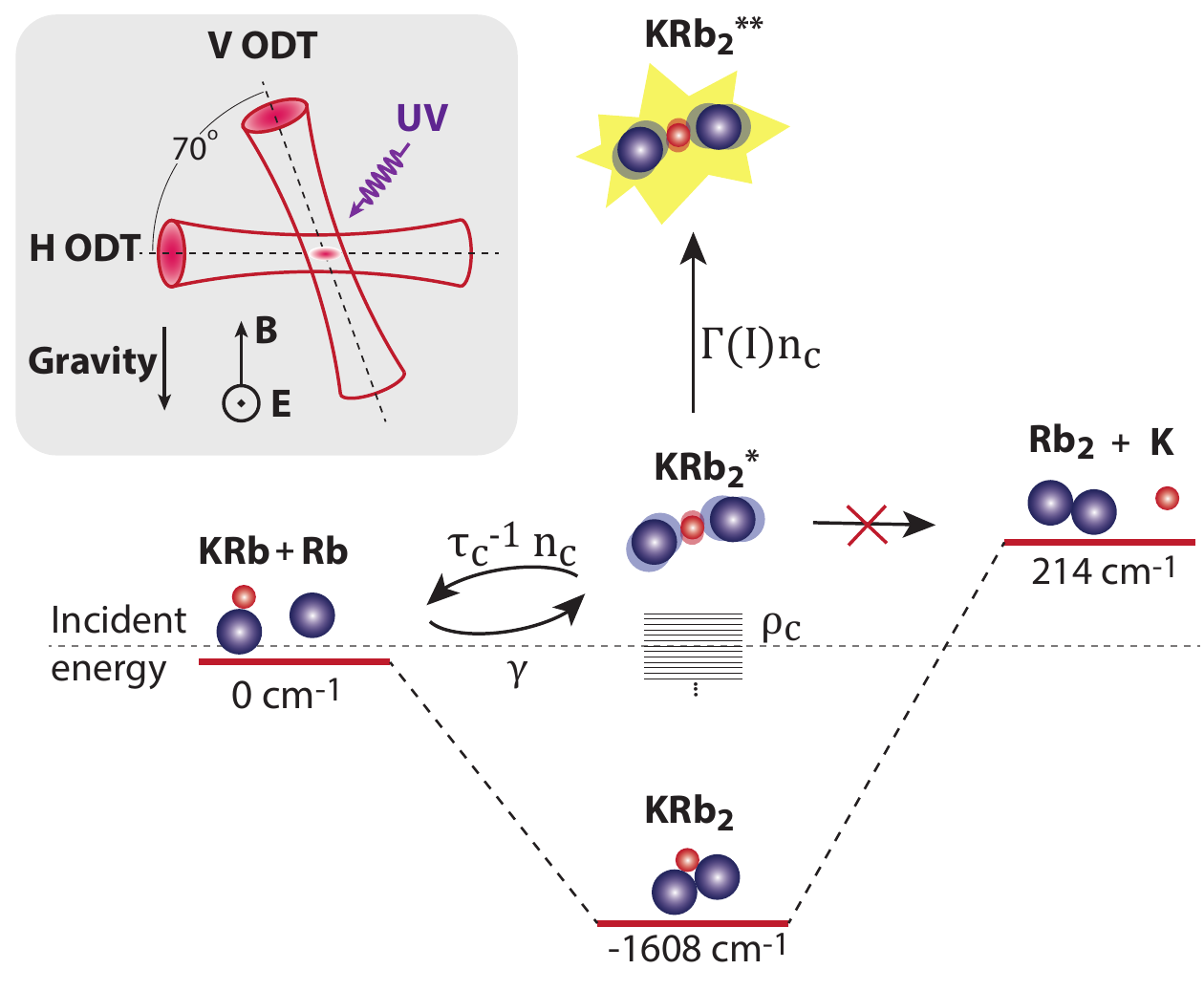}
\caption{Ultracold atom-molecule collisions in an optical dipole trap (ODT). (Inset) A diagram of the  ODT configuration used to confine the mixture of KRb molecules and Rb atoms. The `H' and `V' ODTs are formed from $1064$ nm Gaussian beams with $1/e^2$ diameters of $70$ and $200$ $\mu$m, respectively. The purple arrow represents the UV ionization laser pulse, which has a duration of $7$~ns, used for photoionization of KRb$_{2}^{*}$ complexes. Electric ($E$) and magnetic ($B$) fields are applied to extract the ions, and to maintain nuclear spin quantization, respectively. (b) Schematic illustration of the potential energy surface for collisions between KRb and Rb. The incident energy of one free Rb atom and one free KRb molecule in its rovibronic ground state is defined as the zero of energy. Because the Rb$_{2}\,+\,$K reaction channel is energetically forbidden at ultralow temperatures, the collision must proceed through one of two pathways: KRb$\,+\,$Rb$\,\leftrightarrow\,$KRb$_{2}^{*}$ or KRb$\,+\,$Rb$\,\rightarrow\,$KRb$_{2}^{*}\rightarrow\,$KRb$_{2}^{**}$, where KRb$_{2}^{**}$ is an electronically excited state of the complex. The formation, dissociation, and photo-excitation rates of KRb$_{2}^{*}$, as defined in Eq.~\ref{eq2}, are labeled as $\gamma$, $\tau_{c}^{-1}n_{c}$, and $\Gamma(I_{tot})n_{c}$, respectively, and the density of states (DOS) of the complex near the incident energy is written as $\rho_{c}$. The ground state energies of KRb~\cite{ni2008high} and Rb$_{2}$~\cite{seto2000direct} are obtained from spectroscopic data, whereas that of KRb$_{2}$ is calculated~\cite{hu2019direct}.}
\label{fig1}
\end{figure}

We begin each experiment by creating a gaseous mixture of rotational, vibrational, and electronic (rovibronic) ground-state, fermionic KRb molecules and electronic ground-state Rb atoms at a temperature of $480$ nK. We confine the mixture in a crossed-beam optical dipole trap (ODT), which is formed from two $1064$ nm laser beams (`H' and `V' in Fig.~\ref{fig1}), with a typical  molecular density of $5\times10^{11}$ cm$^{-3}$ and an experimentally variable atomic density of up to $1.6\times10^{12}$ cm$^{-3}$. Using a procedure described in previous works~\cite{ni2008high,OspelkausHyperfine2010, liu2020probing}, the molecular sample is prepared in a single hyperfine state, $\kt{m_{I}^{\text{K}}=-4,m_{I}^{\text{Rb}}=1/2}$, where $m_{I}$ represents the nuclear spin projection onto the quantization magnetic field (Fig.~\ref{fig1} inset). The atoms, on the other hand, occupy the lowest energy hyperfine level, $\kt{F=1,m_{F}=1}$. Because the chemical reaction, $\text{KRb}+\text{Rb}\rightarrow\text{Rb}_{2}+\text{K}$, is endothermic by approximately $214$ cm$^{-1}$ (Fig.~\ref{fig1}), this atom-molecule mixture is chemically stable. Additionally, although the molecules are not prepared in their lowest energy hyperfine state~\cite{OspelkausHyperfine2010,Aldegunde2008}, inelastic collisions that flip the molecular nuclear spins are expected to be suppressed if the total angular momentum is conserved~\cite{Supplement}. In this case, the lifetime of the mixture should be comparable to the lifetime of the bare molecular gas without atoms, which has a measured half-life of $150(8)$ ms.

\begin{figure}[t]
\centering
\includegraphics[width = 0.45\textwidth]{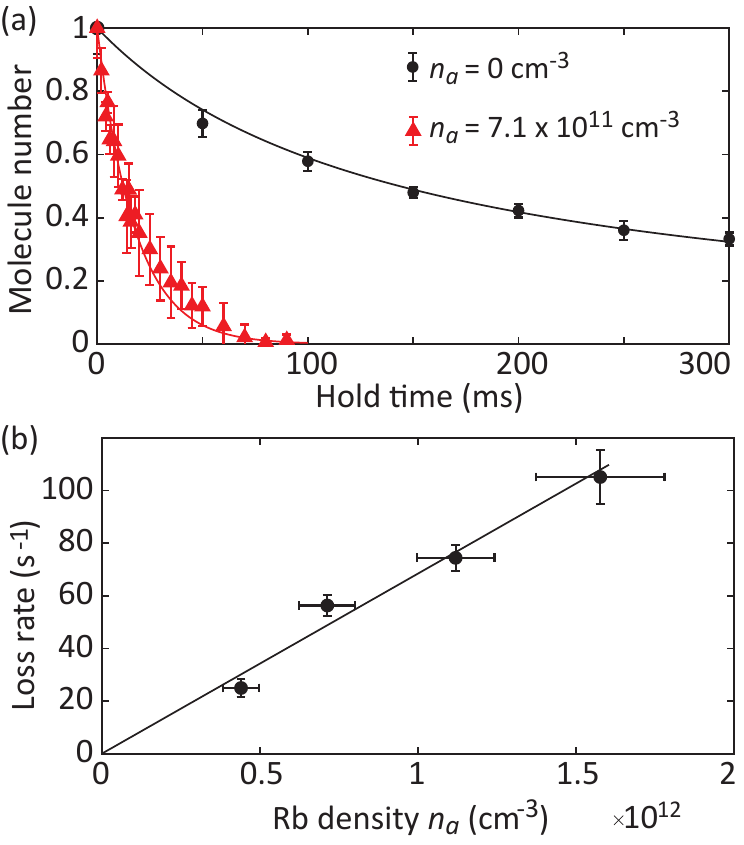}
\caption{Two-body collisional loss in an ultracold mixture of KRb molecules and Rb atoms. (a) Normalized molecule number versus hold time for a pure molecular sample (black circles) and for a representative atom-molecule mixture (red triangles). Each data point is an average of 10 repetitions. The black solid line is a fit to the two-body decay arising from molecule-molecule collisions~\cite{ospelkaus2010quantum}, and the red solid line is a fit using the function $N_{m}(t)=\text{exp}(-k_{m,a}n_{a}t)$. (b) Dependence of the molecular loss rate on the atomic density. A linear fit (black solid line) yields $k_{m,a}=6.8(7)\times10^{-11}$ cm$^3$s$^{-1}$, which is consistent with the predicted universal loss rate~\cite{LiUniversalLoss2019}. All data shown in (a) and (b) are taken with $B=542$ G in a continuously operated ODT of intensity $I_{tot}=11.3$ kW/cm$^2$.}
\label{fig2}
\end{figure}

Upon creation of the atom-molecule mixture, however, we find that the molecule lifetime is reduced significantly relative to the pure molecular sample (Fig.~\ref{fig2}(a)). To quantify this, we prepare a mixture with a desired initial Rb density at a magnetic field of $B=542$ G, and hold it in the ODT for a variable amount of time. We then quickly remove the excess Rb atoms, and perform absorption imaging on the remaining molecules~\cite{Supplement}. To extract the inelastic atom-molecule collision rate, we model the molecular loss via the equation, $dN_{m}/dt=-k_{m,a}N_{m}n_{a}$, where $N_{m}$ denotes the molecule number, $k_{m,a}$ is the two-body loss rate coefficient, and $n_{a}$ represents the atomic density. Because atom-molecule collisions are the dominant loss mechanism, we neglect the background loss associated with reactive molecule-molecule collisions~\cite{ospelkaus2010quantum}. In addition, we assume that $N_{m}$ decays exponentially. This allows us to extract the molecular loss rate for different values of $n_{a}$  and  obtain the two-body loss rate coefficient, $k_{m,a}=6.8(7)\times10^{-11}$ cm$^3$s$^{-1}$, as shown in Fig.~\ref{fig2}(b). This value agrees with the predicted universal loss rate for $s$-wave collisions between KRb and Rb, $k_{m,a}=7.0\times10^{-11}$ cm$^3$s$^{-1}$~\cite{LiUniversalLoss2019}, indicating that atoms and molecules which reach the short-range part of the interaction potential are lost with unit probability.

To gain insight into the source of this universal loss, we probe the atom-molecule collision complex, KRb$_{2}^{*}$, directly by combining single-photon ionization with ion time-of-flight mass spectrometry~\cite{hu2019direct,liu2020probing}. Specifically, after initially preparing the gaseous mixture at a magnetic field of $B=30$ G, we investigate the possibility of trap-light-induced photo-excitation of KRb$_{2}^{*}$ intermediates using techniques similar to those previously demonstrated in Ref.~\cite{liu2020photo}. That is, we apply a $1.5$ kHz square-wave modulation to the intensity of both the H and V ODTs with a $25\%$ duty cycle. By setting the peak intensity during the ``high" phase of the modulation to a value that is four times that of the continuously operated ODT, we keep the time-averaged intensity of the beams constant. This allows us to probe the KRb$_{2}^{*}$ collision complexes which are formed during the $500$ $\mu$s dark phases of the modulation using a pulsed ultraviolet (UV) ionization laser that operates at $355$ nm. The photoionized complexes are then accelerated by a $17$ V/cm electric field onto an ion detector for counting. Ion signals are recorded from each modulation period until the sample is depleted (${\sim}\,1$ s). The total number of KRb$_{2}^{+}$ ions counted during an experimental cycle acts as a proxy for the KRb$_{2}^{*}$ population at the instance of time within the modulation period where the UV laser pulse occurs. As the pulse repetition frequency is synchronized with the ODT modulation, we can control the relative delay between the two to probe the KRb$_{2}^{*}$ population at different points during the modulation period.

\begin{figure}[t]
\centering
\includegraphics[width = 0.44\textwidth]{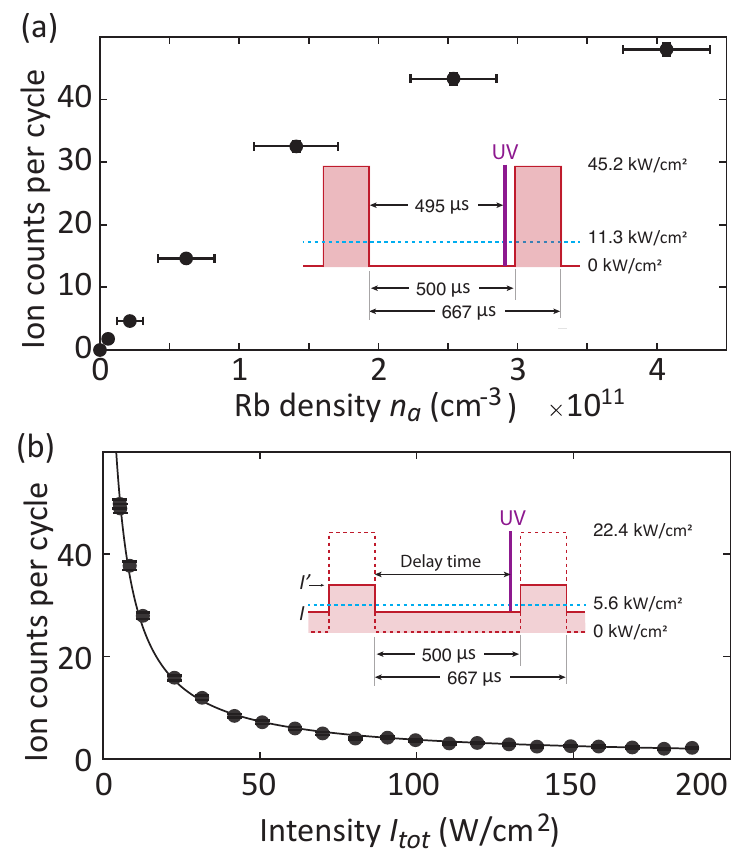}
\caption{Photo-excitation loss of KRb$_{2}^{*}$ collision complexes in a $1064$ nm optical trap. (a) Steady-state KRb$_{2}^{+}$ ion counts measured for different initial atom number densities, $n_{a}$, and normalized by the number of experimental cycles ($\sim100$ for each data point). The inset shows the timing diagram used for the measurement. The red curve is the total ODT intensity and the blue dashed line is its time average. (b) Steady-state KRb$_{2}^{+}$ ion counts measured at different total ODT intensities with $n_{a}=4.1(3)\times10^{11}$ cm$^{-3}$, and normalized by the number of experimental cycles ($\sim50$ for each data point). The solid line is a fit to the data using Eq.~\ref{eq4}. The inset shows the timing diagram used for the measurement. The red curve is the intensity of the V ODT, where the intensity during the ``high'' phase of the modulation period is $I'$, and that during the ``low'' phase is $I$. To control $I_{tot}$, we vary the V ODT modulation depth while keeping the time-averaged intensity fixed at the continuously operated level, $5.6$ kW/cm$^2$ (blue dashed line). The red dashed line is the intensity profile at full modulation depth. The H ODT, not shown here, follows the V ODT timing, but is always modulated at full depth with a time-averaged intensity equal to its continuously operated level, $5.7$ kW/cm$^2$. All data shown in (a) and (b) are taken with $B=30$ G and $E=17$ V/cm.}
\label{fig3}
\end{figure}

By probing the sample near the end of the dark phase of each modulation period, we can examine the steady-state complex population in the absence of ODT light. In Fig.~\ref{fig3}(a), we show the measured number of KRb$_{2}^{+}$ ions per experimental cycle for different initial Rb atom number densities. When no atoms are present, we observe no KRb$_{2}^{+}$ counts. As the Rb density is increased from zero, however, so too does the measured number of KRb$_{2}^{+}$ counts, indicating that the KRb$_{2}^{+}$ signal is the result of atom-molecule collisions. At sufficiently high Rb densities, the measured KRb$_{2}^{+}$ counts begin to saturate due to a competition between the finite UV pulse repetition frequency and the decay rate of the sample.

To investigate the effect of the ODT light on the collision complexes, we fix the initial Rb number density at $4.1(3)\times10^{11}$ cm$^{-3}$, which yields the maximal KRb$_{2}^{+}$ signal, and we vary the intensity level of the ``low" phase of the square-wave modulation (Fig.~\ref{fig3}(b) inset). As shown in Fig.~\ref{fig3}(b), we find that the steady-state KRb$_{2}^{+}$ counts decrease monotonically with increasing optical intensity, which indicates that the complex population is highly suppressed by the presence of $1064$ nm light. Such evidence strongly supports the notion of photo-induced loss of these transient complexes~\cite{christianen2019photoinduced}.

To quantify this behavior, we model the rate of change of the complex population ($n_{c}$) via the three processes illustrated in Fig.~\ref{fig1}: complex formation through atom-molecule collisions ($\gamma$), dissociation over the time-scale of the complex lifetime ($\tau_{c}$), and an intensity-dependent photo-excitation loss ($\Gamma_{l}(I_{tot})$). The corresponding rate equation is,
\begin{equation}
\dot{n}_{c}(t)=\gamma-\tau_{c}^{-1}n_{c}(t)-\Gamma_{l}(I_{tot})n_{c}(t).
\label{eq2}
\end{equation}
Taking $\Gamma_{l}(I_{tot})=\beta_{1}I_{tot}+\beta_{2}I_{tot}^2$, where $\beta_{1}$ describes single-photon excitation of the complex and $\beta_{2}$ represents second-order photo-excitation processes that are empirically found to be important at large intensities~\cite{liu2020photo}, the steady-state solution to this equation is given by $n_{c}=\gamma\tau_{c}/(1+\beta_{1}\tau_{c}I_{tot}+\beta_{2}\tau_{c}I_{tot}^2)$. As the measured KRb$_{2}^{+}$ ion counts in Fig.~\ref{fig3}(b) directly reflect the steady-state complex population, their dependence on the total $1064$ nm light intensity can be expressed as,
\begin{equation}
N_{\text{KRb}_{2}^{+}}(I_{tot})=\frac{A}{1+B_{1}I_{tot}+B_{2}I_{tot}^2},
\label{eq4}
\end{equation}
where $A$ describes the number of complex ions obtained at zero intensity, and $B_{1,2}\equiv\beta_{1,2}\tau_{c}$. Fitting the data in Fig.~\ref{fig3}(b) using Eq.~\ref{eq4}, we obtain  $B_{1}=0.26(5)$ $($W/cm$^{2})^{-1}$ and $B_{2}=0.0008(3)$ $($W/cm$^{2})^{-2}$. Because the term $B_{1}I_{tot}+B_{2}I_{tot}^2$ physically represents the branching ratio between the rate of photo-excitation by the ODT and the rate of dissociation back into KRb molecules and Rb atoms, we find that the KRb$_{2}^{*}$ intermediates are $10^5$ times more likely to undergo photo-excitation at the typical total intensity used for the continuously operated crossed ODT, $11.3$ kW/cm$^2$, than they are to dissociate. Therefore, under standard operating conditions for the crossed ODT, any atoms and molecules which reach the short-range part of the interaction potential are lost via trap-induced photo-excitation with near-unit probability.

While this may account for the universal loss observed in the atom-molecule mixture, this measurement alone cannot determine whether the sensitivity of the intermediates to the ODT light arises from large photo-excitation rates, a long complex lifetime, or some combination of the two. In order to disentangle these effects, we utilize the optical excitation of the complex to induce dynamics in the KRb$_{2}^{*}$ population. Specifically, by exposing the sample to enough ODT intensity, we can deplete the KRb$_{2}^{*}$ population and establish a zero-of-time~\cite{liu2020photo}. By then rapidly reducing the intensity, we cause the complex population to grow towards a new steady state. In the limit where $\gamma$ is effectively constant over the short timescales associated with such dynamics, the solution to Eq.~\ref{eq2} for times immediately following the intensity change is,
\begin{equation}
n_{c}(t)=\frac{\gamma}{\tau_{c}^{-1}+\Gamma_{l}(I_{tot})}\left(1-e^{-\left(\tau_{c}^{-1}+\Gamma_{l}(I_{tot})\right)t}\right),
\label{eq5}
\end{equation}
where $I_{tot}$ here is the final $1064$ nm intensity. For low intensities, where the second-order contribution to $\Gamma_{l}(I_{tot})$ is negligible, the characteristic growth rate, $R$, of the population can be written as $R=\tau_{c}^{-1}+\beta_{1}I_{tot}$, which scales linearly with $I_{tot}$. By extracting $R$ for different values of $I_{tot}$, we can measure this linear dependence to obtain both $\beta_{1}$ and $\tau_{c}$.

\begin{figure}[t]
\centering
\includegraphics[width = 0.45\textwidth]{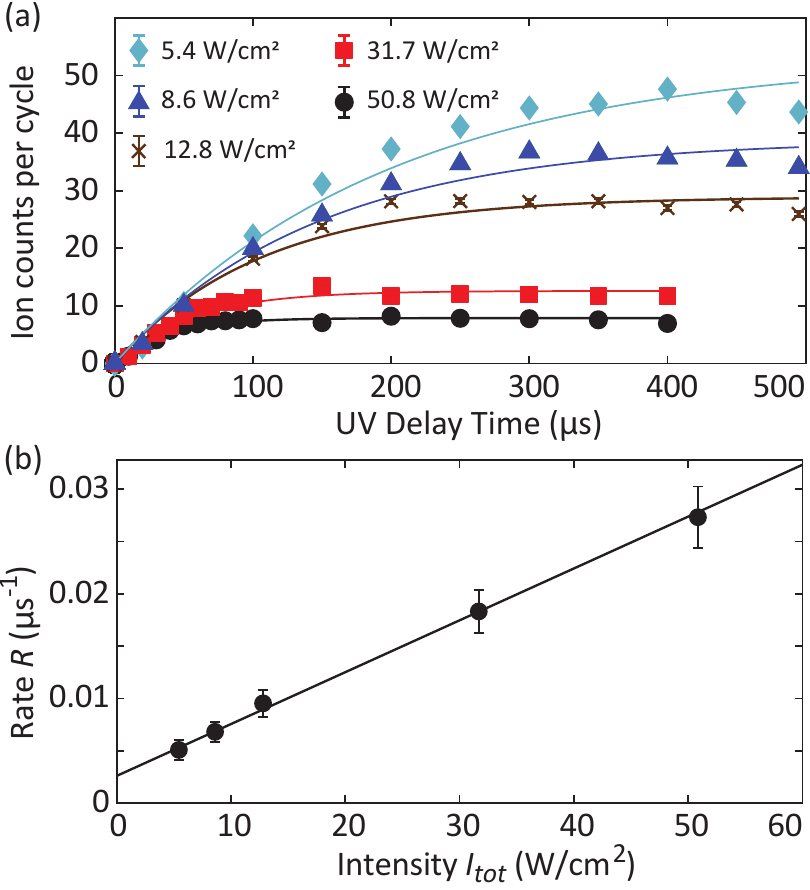}
\caption{Lifetime of the KRb$_{2}^{*}$ collision complex. (a) Time evolution of the KRb$_{2}^{*}$ population (measured via the KRb$_{2}^{+}$ ion signal) after a rapid change in the $1064$ nm intensity, for different values of $I_{tot}$. Data is normalized by the corresponding number of experimental cycles ($\sim100$ for each data point). The time-averaged value of the total intensity is fixed at $11.3$ kW/cm$^2$ across all data sets. The solid lines are fits using the function $A\left(1-e^{-Rt}\right)$. (b) Characteristic growth rate, $R$, of the complex population, as obtained from the fits in (a), versus the corresponding total ODT intensity. The solid line is a linear fit to the data. All data shown in (a) and (b) are taken with $B=30$ G and $E=17$ V/cm.}
\label{fig4}
\end{figure}

To do this, we apply the same modulation scheme shown in the inset to Fig.~\ref{fig3}(b), but we vary the delay time between the UV pulse and the off-edge of the intensity modulation. In Fig.~\ref{fig4}(a), we show the measured KRb$_{2}^{+}$ counts versus this delay time for different values of $I_{tot}$. We fit these curves using the function, $N_{\text{KRb}_{2}^{+}}=A\left(1-e^{-Rt}\right)$ (derived from Eq.~\ref{eq5}), to extract $R$, which is plotted versus $I_{tot}$ in Fig.~\ref{fig4}(b). As expected, this rate increases linearly with increasing values of $I_{tot}$. We therefore fit the data using a linear function, which yields $\tau_{c}=0.39(6)$ ms and $\beta_{1}=0.50(3)$ $\mu$s$^{-1}($kW/cm$^{2})^{-1}$. The corresponding value of $B_{1}=0.20(3)$ $($W/cm$^{2})^{-1}$ agrees with that obtained from the data in Fig.~\ref{fig3}(b) to within the measurement uncertainty. We can further determine the value of $\beta_{2}$ using both $\tau_{c}$ and the value of $B_{2}=\beta_{2}\tau_{c}$ extracted from the data in Fig.~\ref{fig3}(b), which yields $\beta_{2}=2.1(8)$ $\mu$s$^{-1}($kW/cm$^{2})^{-1}$, similar to what was found previously for the photo-excitation of K$_2$Rb$_2^*$~\cite{liu2020photo}.

We therefore conclude that the observed sensitivity of KRb$_{2}^{*}$ to $1064$ nm light is predominantly the result of an exceptionally long lifetime. Theoretical estimates of this lifetime have previously been made using RRKM statistical theory~\cite{levine2009molecular}, which assumes that the complex ergodically explores the reaction phase space before it dissociates. The corresponding RRKM complex lifetime is given by $\tau_{c}=2\pi\hbar\rho_{c}/\mathcal{N}$, where $\rho_{c}$ represents the density of states (DOS) of the complex (Fig.~\ref{fig1}), and $\mathcal{N}$ denotes the number of available dissociation channels. Previous calculations of this lifetime for KRb$_{2}^{*}$ are $270$ ns~\cite{mayle2012statistical} and $1$ ns~\cite{PrivateCommunication,ChristianenDOS2019}, which differ from our results by factors of ${\sim}10^3$ and ${\sim}10^5$, respectively. As similar calculations have previously shown good agreement with measurements for the cases of K$_{2}$Rb$_{2}^{*}$ complexes formed through reactive $^{40}$K$^{87}$Rb -$\,^{40}$K$^{87}$Rb collisions~\cite{liu2020photo}, and Rb$_{2}$Cs$_{2}^{*}$ complexes formed through non-reactive $^{87}$Rb$^{133}$Cs -$\,^{87}$Rb$^{133}$Cs collisions~\cite{gregory2020loss}, such a drastic underestimation by the RRKM predictions is surprising.

One possible explanation for this is the potential absence of angular momentum conservation in the atom-molecule collisions studied here. Such a phenomenon could arise either from the presence of external fields in the experiment, or from internal couplings which enable hyperfine transitions within the complex. While the quoted RRKM lifetimes assume that the total angular momentum is conserved, the breakdown of this assumption could increase the DOS of the complex, and correspondingly the RRKM lifetime, by several orders of magnitude~\cite{ChristianenDOS2019}. Unfortunately, calculations of the critical electric and magnetic field strengths required to increase the DOS in this way are not currently available. To examine this effect experimentally, we have measured the KRb$_{2}^{*}$ lifetime for different electric and magnetic fields in the ranges of $17-343$ V/cm and $30-300$ G, respectively, but we observe no significant variation~\cite{Supplement}. As for the potential existence of couplings within the complex which allow for hyperfine transitions, the results of previous studies suggest that, at least in the case of ultracold collisions between $^{40}$K$^{87}$Rb molecules, both nuclear spins and the total angular momentum are conserved throughout the chemical reaction, $2$KRb$\,\rightarrow\,$K$_{2}$Rb$_{2}^{*}\rightarrow\,$K$_{2}+$Rb$_{2}$~\cite{hu2020product,liu2020precision}. While similar behavior might also be expected for collisions between KRb molecules and Rb atoms, further investigations are needed.

Another potential reason for the discrepancy between the RRKM predictions and the observed lifetime is that the RRKM statistical theory simply may not provide an accurate description of the collision dynamics for this system. This could occur, for example, if the relevant DOS of the triatomic KRb$_{2}^{*}$ collision complex is small compared to the initial sample temperature of $480$ nK, so that few scattering resonances lie within the range of accessible energies. In this case, the statistical assumption that the complex can ergodically explore many different narrow resonances in phase space before dissociating would not apply, and so one would not necessarily expect the RRKM lifetime to be quantitatively accurate.

In summary, we have observed universal two-body collisional loss in an ultracold mixture of $^{40}$K$^{87}$Rb molecules and $^{87}$Rb atoms, which we attribute to photo-excitation of the KRb$_{2}^{*}$ collision complexes by the $1064$ nm lasers used to trap the sample. Through the direct detection of these complexes, we measure their lifetime, as well as the intensity-dependent photo-excitation rate for $1064$ nm light. The measured complex lifetime deviates from the most recent predictions by up to five orders of magnitude, and therefore demands further theoretical investigations. The long lifetime we observe for the KRb$_{2}^{*}$ intermediates may also shed light on recent results from studies of collisional loss in chemically stable, ultracold gases of NaK and NaRb molecules~\cite{gersema2021probing, bause2021collisions}. In those systems, although direct detection of intermediates was not possible, experimental observations suggest that the corresponding complex lifetimes may be several orders of magnitude longer than previously predicted~\cite{ChristianenDOS2019,christianen2019photoinduced}. Thus, while our results provide a direct demonstration that these ultracold collision complexes do, in fact, live much longer than previously thought, new theoretical insight is required to reveal the physical mechanisms underlying this phenomenon. Such insight will help to guide the development of a new understanding of ultracold molecular scattering, which should ultimately apply to both atom-molecule and molecule-molecule collisions. This is crucial for the production of molecular gases with higher phase-space densities, as well as the realization of the many exciting research opportunities in quantum simulation, quantum information, and precision measurements~\cite{baranov2012condensed, demille2002quantum,ni2018dipolar,Safronova2018Review} such systems promise.

\begin{acknowledgments}
We thank Svetlana Kotochigova, John L. Bohn, Tijs Karman, and Jeremy Hutson for insightful discussions. This work was supported by the DOE Young Investigator Program (DE-SC0019020) and the David and Lucile Packard Foundation. M.A.N. was supported by the Arnold O. Beckman Postdoctoral Fellowship in Chemical Instrumentation.
\end{acknowledgments}

\bibliographystyle{apsrev4-1}

\begin{thebibliography}{49}%
\makeatletter
\providecommand \@ifxundefined [1]{%
 \@ifx{#1\undefined}
}%
\providecommand \@ifnum [1]{%
 \ifnum #1\expandafter \@firstoftwo
 \else \expandafter \@secondoftwo
 \fi
}%
\providecommand \@ifx [1]{%
 \ifx #1\expandafter \@firstoftwo
 \else \expandafter \@secondoftwo
 \fi
}%
\providecommand \natexlab [1]{#1}%
\providecommand \enquote  [1]{``#1''}%
\providecommand \bibnamefont  [1]{#1}%
\providecommand \bibfnamefont [1]{#1}%
\providecommand \citenamefont [1]{#1}%
\providecommand \href@noop [0]{\@secondoftwo}%
\providecommand \href [0]{\begingroup \@sanitize@url \@href}%
\providecommand \@href[1]{\@@startlink{#1}\@@href}%
\providecommand \@@href[1]{\endgroup#1\@@endlink}%
\providecommand \@sanitize@url [0]{\catcode `\\12\catcode `\$12\catcode
  `\&12\catcode `\#12\catcode `\^12\catcode `\_12\catcode `\%12\relax}%
\providecommand \@@startlink[1]{}%
\providecommand \@@endlink[0]{}%
\providecommand \url  [0]{\begingroup\@sanitize@url \@url }%
\providecommand \@url [1]{\endgroup\@href {#1}{\urlprefix }}%
\providecommand \urlprefix  [0]{URL }%
\providecommand \Eprint [0]{\href }%
\providecommand \doibase [0]{http://dx.doi.org/}%
\providecommand \selectlanguage [0]{\@gobble}%
\providecommand \bibinfo  [0]{\@secondoftwo}%
\providecommand \bibfield  [0]{\@secondoftwo}%
\providecommand \translation [1]{[#1]}%
\providecommand \BibitemOpen [0]{}%
\providecommand \bibitemStop [0]{}%
\providecommand \bibitemNoStop [0]{.\EOS\space}%
\providecommand \EOS [0]{\spacefactor3000\relax}%
\providecommand \BibitemShut  [1]{\csname bibitem#1\endcsname}%
\let\auto@bib@innerbib\@empty
%</preamble>
\bibitem [{\citenamefont {Quemener}\ and\ \citenamefont
  {Julienne}(2012)}]{quemener2012ultracold}%
  \BibitemOpen
  \bibfield  {author} {\bibinfo {author} {\bibfnamefont {G.}~\bibnamefont
  {Quemener}}\ and\ \bibinfo {author} {\bibfnamefont {P.~S.}\ \bibnamefont
  {Julienne}},\ }\href@noop {} {\bibfield  {journal} {\bibinfo  {journal}
  {Chem. Rev.}\ }\textbf {\bibinfo {volume} {112}},\ \bibinfo {pages} {4949}
  (\bibinfo {year} {2012})}\BibitemShut {NoStop}%
\bibitem [{\citenamefont {Jankunas}\ and\ \citenamefont
  {Osterwalder}(2015)}]{Jankunas2015}%
  \BibitemOpen
  \bibfield  {author} {\bibinfo {author} {\bibfnamefont {J.}~\bibnamefont
  {Jankunas}}\ and\ \bibinfo {author} {\bibfnamefont {A.}~\bibnamefont
  {Osterwalder}},\ }\href@noop {} {\bibfield  {journal} {\bibinfo  {journal}
  {Annu. Rev. Phys. Chem.}\ }\textbf {\bibinfo {volume} {66}},\ \bibinfo
  {pages} {241} (\bibinfo {year} {2015})}\BibitemShut {NoStop}%
\bibitem [{\citenamefont {Bohn}\ \emph {et~al.}(2017)\citenamefont {Bohn},
  \citenamefont {Rey},\ and\ \citenamefont {Ye}}]{Bohn2017Review}%
  \BibitemOpen
  \bibfield  {author} {\bibinfo {author} {\bibfnamefont {J.~L.}\ \bibnamefont
  {Bohn}}, \bibinfo {author} {\bibfnamefont {A.~M.}\ \bibnamefont {Rey}}, \
  and\ \bibinfo {author} {\bibfnamefont {J.}~\bibnamefont {Ye}},\ }\href
  {\doibase 10.1126/science.aam6299} {\bibfield  {journal} {\bibinfo  {journal}
  {Science}\ }\textbf {\bibinfo {volume} {357}},\ \bibinfo {pages} {1002}
  (\bibinfo {year} {2017})}\BibitemShut {NoStop}%
\bibitem [{\citenamefont {Yang}(2007)}]{yang2007state}%
  \BibitemOpen
  \bibfield  {author} {\bibinfo {author} {\bibfnamefont {X.}~\bibnamefont
  {Yang}},\ }\href@noop {} {\bibfield  {journal} {\bibinfo  {journal} {Annu.
  Rev. Phys. Chem.}\ }\textbf {\bibinfo {volume} {58}},\ \bibinfo {pages} {433}
  (\bibinfo {year} {2007})}\BibitemShut {NoStop}%
\bibitem [{\citenamefont {Balakrishnan}(2016)}]{balakrishnan2016perspective}%
  \BibitemOpen
  \bibfield  {author} {\bibinfo {author} {\bibfnamefont {N.}~\bibnamefont
  {Balakrishnan}},\ }\href@noop {} {\bibfield  {journal} {\bibinfo  {journal}
  {J. Chem. Phys.}\ }\textbf {\bibinfo {volume} {145}},\ \bibinfo {pages}
  {150901} (\bibinfo {year} {2016})}\BibitemShut {NoStop}%
\bibitem [{\citenamefont {Dulieu}\ and\ \citenamefont
  {Osterwalder}(2018)}]{Dulieu2018Book}%
  \BibitemOpen
  \bibinfo {editor} {\bibfnamefont {O.}~\bibnamefont {Dulieu}}\ and\ \bibinfo
  {editor} {\bibfnamefont {A.}~\bibnamefont {Osterwalder}},\ eds.,\ \href@noop
  {} {\emph {\bibinfo {title} {Cold Chemistry: Molecular Scattering and
  Reactivity Near Absolute Zero}}}\ (\bibinfo  {publisher} {The Royal Society
  of Chemistry},\ \bibinfo {year} {2018})\BibitemShut {NoStop}%
\bibitem [{\citenamefont {Toscano}\ \emph {et~al.}(2020)\citenamefont
  {Toscano}, \citenamefont {Lewandowski},\ and\ \citenamefont
  {Heazlewood}}]{toscano2020cold}%
  \BibitemOpen
  \bibfield  {author} {\bibinfo {author} {\bibfnamefont {J.}~\bibnamefont
  {Toscano}}, \bibinfo {author} {\bibfnamefont {H.}~\bibnamefont
  {Lewandowski}}, \ and\ \bibinfo {author} {\bibfnamefont {B.~R.}\ \bibnamefont
  {Heazlewood}},\ }\href@noop {} {\bibfield  {journal} {\bibinfo  {journal}
  {Phys. Chem. Chem. Phys.}\ }\textbf {\bibinfo {volume} {22}},\ \bibinfo
  {pages} {9180} (\bibinfo {year} {2020})}\BibitemShut {NoStop}%
\bibitem [{\citenamefont {Liu}\ \emph {et~al.}(2020{\natexlab{a}})\citenamefont
  {Liu}, \citenamefont {Grimes}, \citenamefont {Hu},\ and\ \citenamefont
  {Ni}}]{liu2020probing}%
  \BibitemOpen
  \bibfield  {author} {\bibinfo {author} {\bibfnamefont {Y.}~\bibnamefont
  {Liu}}, \bibinfo {author} {\bibfnamefont {D.~D.}\ \bibnamefont {Grimes}},
  \bibinfo {author} {\bibfnamefont {M.-G.}\ \bibnamefont {Hu}}, \ and\ \bibinfo
  {author} {\bibfnamefont {K.-K.}\ \bibnamefont {Ni}},\ }\href@noop {}
  {\bibfield  {journal} {\bibinfo  {journal} {Physical Chemistry Chemical
  Physics}\ }\textbf {\bibinfo {volume} {22}},\ \bibinfo {pages} {4861}
  (\bibinfo {year} {2020}{\natexlab{a}})}\BibitemShut {NoStop}%
\bibitem [{\citenamefont {Klein}\ \emph {et~al.}(2017)\citenamefont {Klein},
  \citenamefont {Shagam}, \citenamefont {Skomorowski}, \citenamefont
  {{\.Z}uchowski}, \citenamefont {Pawlak}, \citenamefont {Janssen},
  \citenamefont {Moiseyev}, \citenamefont {van~de Meerakker}, \citenamefont
  {van~der Avoird}, \citenamefont {Koch},\ and\ \citenamefont
  {Narevicius}}]{klein2017directly}%
  \BibitemOpen
  \bibfield  {author} {\bibinfo {author} {\bibfnamefont {A.}~\bibnamefont
  {Klein}}, \bibinfo {author} {\bibfnamefont {Y.}~\bibnamefont {Shagam}},
  \bibinfo {author} {\bibfnamefont {W.}~\bibnamefont {Skomorowski}}, \bibinfo
  {author} {\bibfnamefont {P.~S.}\ \bibnamefont {{\.Z}uchowski}}, \bibinfo
  {author} {\bibfnamefont {M.}~\bibnamefont {Pawlak}}, \bibinfo {author}
  {\bibfnamefont {L.~M.}\ \bibnamefont {Janssen}}, \bibinfo {author}
  {\bibfnamefont {N.}~\bibnamefont {Moiseyev}}, \bibinfo {author}
  {\bibfnamefont {S.~Y.}\ \bibnamefont {van~de Meerakker}}, \bibinfo {author}
  {\bibfnamefont {A.}~\bibnamefont {van~der Avoird}}, \bibinfo {author}
  {\bibfnamefont {C.~P.}\ \bibnamefont {Koch}}, \ and\ \bibinfo {author}
  {\bibfnamefont {E.}~\bibnamefont {Narevicius}},\ }\href@noop {} {\bibfield
  {journal} {\bibinfo  {journal} {Nat. Phys.}\ }\textbf {\bibinfo {volume}
  {13}},\ \bibinfo {pages} {35} (\bibinfo {year} {2017})}\BibitemShut {NoStop}%
\bibitem [{\citenamefont {Yang}\ \emph {et~al.}(2019)\citenamefont {Yang},
  \citenamefont {Zhang}, \citenamefont {Liu}, \citenamefont {Liu},
  \citenamefont {Nan}, \citenamefont {Zhao},\ and\ \citenamefont
  {Pan}}]{yang2019observation}%
  \BibitemOpen
  \bibfield  {author} {\bibinfo {author} {\bibfnamefont {H.}~\bibnamefont
  {Yang}}, \bibinfo {author} {\bibfnamefont {D.-C.}\ \bibnamefont {Zhang}},
  \bibinfo {author} {\bibfnamefont {L.}~\bibnamefont {Liu}}, \bibinfo {author}
  {\bibfnamefont {Y.-X.}\ \bibnamefont {Liu}}, \bibinfo {author} {\bibfnamefont
  {J.}~\bibnamefont {Nan}}, \bibinfo {author} {\bibfnamefont {B.}~\bibnamefont
  {Zhao}}, \ and\ \bibinfo {author} {\bibfnamefont {J.-W.}\ \bibnamefont
  {Pan}},\ }\href@noop {} {\bibfield  {journal} {\bibinfo  {journal} {Science}\
  }\textbf {\bibinfo {volume} {363}},\ \bibinfo {pages} {261} (\bibinfo {year}
  {2019})}\BibitemShut {NoStop}%
\bibitem [{\citenamefont {de~Jongh}\ \emph {et~al.}(2020)\citenamefont
  {de~Jongh}, \citenamefont {Besemer}, \citenamefont {Shuai}, \citenamefont
  {Karman}, \citenamefont {van~der Avoird}, \citenamefont {Groenenboom},\ and\
  \citenamefont {van~de Meerakker}}]{de2020imaging}%
  \BibitemOpen
  \bibfield  {author} {\bibinfo {author} {\bibfnamefont {T.}~\bibnamefont
  {de~Jongh}}, \bibinfo {author} {\bibfnamefont {M.}~\bibnamefont {Besemer}},
  \bibinfo {author} {\bibfnamefont {Q.}~\bibnamefont {Shuai}}, \bibinfo
  {author} {\bibfnamefont {T.}~\bibnamefont {Karman}}, \bibinfo {author}
  {\bibfnamefont {A.}~\bibnamefont {van~der Avoird}}, \bibinfo {author}
  {\bibfnamefont {G.~C.}\ \bibnamefont {Groenenboom}}, \ and\ \bibinfo {author}
  {\bibfnamefont {S.~Y.~T.}\ \bibnamefont {van~de Meerakker}},\ }\href@noop {}
  {\bibfield  {journal} {\bibinfo  {journal} {Science}\ }\textbf {\bibinfo
  {volume} {368}},\ \bibinfo {pages} {626} (\bibinfo {year}
  {2020})}\BibitemShut {NoStop}%
\bibitem [{\citenamefont {Ospelkaus}\ \emph
  {et~al.}(2010{\natexlab{a}})\citenamefont {Ospelkaus}, \citenamefont {Ni},
  \citenamefont {Wang}, \citenamefont {De~Miranda}, \citenamefont {Neyenhuis},
  \citenamefont {Qu{\'e}m{\'e}ner}, \citenamefont {Julienne}, \citenamefont
  {Bohn}, \citenamefont {Jin},\ and\ \citenamefont
  {Ye}}]{ospelkaus2010quantum}%
  \BibitemOpen
  \bibfield  {author} {\bibinfo {author} {\bibfnamefont {S.}~\bibnamefont
  {Ospelkaus}}, \bibinfo {author} {\bibfnamefont {K.-K.}\ \bibnamefont {Ni}},
  \bibinfo {author} {\bibfnamefont {D.}~\bibnamefont {Wang}}, \bibinfo {author}
  {\bibfnamefont {M.}~\bibnamefont {De~Miranda}}, \bibinfo {author}
  {\bibfnamefont {B.}~\bibnamefont {Neyenhuis}}, \bibinfo {author}
  {\bibfnamefont {G.}~\bibnamefont {Qu{\'e}m{\'e}ner}}, \bibinfo {author}
  {\bibfnamefont {P.}~\bibnamefont {Julienne}}, \bibinfo {author}
  {\bibfnamefont {J.}~\bibnamefont {Bohn}}, \bibinfo {author} {\bibfnamefont
  {D.}~\bibnamefont {Jin}}, \ and\ \bibinfo {author} {\bibfnamefont
  {J.}~\bibnamefont {Ye}},\ }\href@noop {} {\bibfield  {journal} {\bibinfo
  {journal} {Science}\ }\textbf {\bibinfo {volume} {327}},\ \bibinfo {pages}
  {853} (\bibinfo {year} {2010}{\natexlab{a}})}\BibitemShut {NoStop}%
\bibitem [{\citenamefont {Ni}\ \emph {et~al.}(2010)\citenamefont {Ni},
  \citenamefont {Ospelkaus}, \citenamefont {Wang}, \citenamefont
  {Qu{\'e}m{\'e}ner}, \citenamefont {Neyenhuis}, \citenamefont {De~Miranda},
  \citenamefont {Bohn}, \citenamefont {Ye},\ and\ \citenamefont
  {Jin}}]{ni2010dipolar}%
  \BibitemOpen
  \bibfield  {author} {\bibinfo {author} {\bibfnamefont {K.-K.}\ \bibnamefont
  {Ni}}, \bibinfo {author} {\bibfnamefont {S.}~\bibnamefont {Ospelkaus}},
  \bibinfo {author} {\bibfnamefont {D.}~\bibnamefont {Wang}}, \bibinfo {author}
  {\bibfnamefont {G.}~\bibnamefont {Qu{\'e}m{\'e}ner}}, \bibinfo {author}
  {\bibfnamefont {B.}~\bibnamefont {Neyenhuis}}, \bibinfo {author}
  {\bibfnamefont {M.}~\bibnamefont {De~Miranda}}, \bibinfo {author}
  {\bibfnamefont {J.}~\bibnamefont {Bohn}}, \bibinfo {author} {\bibfnamefont
  {J.}~\bibnamefont {Ye}}, \ and\ \bibinfo {author} {\bibfnamefont
  {D.}~\bibnamefont {Jin}},\ }\href@noop {} {\bibfield  {journal} {\bibinfo
  {journal} {Nature}\ }\textbf {\bibinfo {volume} {464}},\ \bibinfo {pages}
  {1324} (\bibinfo {year} {2010})}\BibitemShut {NoStop}%
\bibitem [{\citenamefont {Hall}\ and\ \citenamefont
  {Willitsch}(2012)}]{hall2012millikelvin}%
  \BibitemOpen
  \bibfield  {author} {\bibinfo {author} {\bibfnamefont {F.~H.~J.}\
  \bibnamefont {Hall}}\ and\ \bibinfo {author} {\bibfnamefont {S.}~\bibnamefont
  {Willitsch}},\ }\href@noop {} {\bibfield  {journal} {\bibinfo  {journal}
  {Phys. Rev. Lett.}\ }\textbf {\bibinfo {volume} {109}},\ \bibinfo {pages}
  {233202} (\bibinfo {year} {2012})}\BibitemShut {NoStop}%
\bibitem [{\citenamefont {Perreault}\ \emph {et~al.}(2017)\citenamefont
  {Perreault}, \citenamefont {Mukherjee},\ and\ \citenamefont
  {Zare}}]{perreault2017quantum}%
  \BibitemOpen
  \bibfield  {author} {\bibinfo {author} {\bibfnamefont {W.~E.}\ \bibnamefont
  {Perreault}}, \bibinfo {author} {\bibfnamefont {N.}~\bibnamefont
  {Mukherjee}}, \ and\ \bibinfo {author} {\bibfnamefont {R.~N.}\ \bibnamefont
  {Zare}},\ }\href@noop {} {\bibfield  {journal} {\bibinfo  {journal}
  {Science}\ }\textbf {\bibinfo {volume} {358}},\ \bibinfo {pages} {356}
  (\bibinfo {year} {2017})}\BibitemShut {NoStop}%
\bibitem [{\citenamefont {Guo}\ \emph {et~al.}(2018)\citenamefont {Guo},
  \citenamefont {Ye}, \citenamefont {He}, \citenamefont
  {Gonz\'alez-Mart\'{\i}nez}, \citenamefont {Vexiau}, \citenamefont
  {Qu\'em\'ener},\ and\ \citenamefont {Wang}}]{guo2018dipolar}%
  \BibitemOpen
  \bibfield  {author} {\bibinfo {author} {\bibfnamefont {M.}~\bibnamefont
  {Guo}}, \bibinfo {author} {\bibfnamefont {X.}~\bibnamefont {Ye}}, \bibinfo
  {author} {\bibfnamefont {J.}~\bibnamefont {He}}, \bibinfo {author}
  {\bibfnamefont {M.~L.}\ \bibnamefont {Gonz\'alez-Mart\'{\i}nez}}, \bibinfo
  {author} {\bibfnamefont {R.}~\bibnamefont {Vexiau}}, \bibinfo {author}
  {\bibfnamefont {G.}~\bibnamefont {Qu\'em\'ener}}, \ and\ \bibinfo {author}
  {\bibfnamefont {D.}~\bibnamefont {Wang}},\ }\href@noop {} {\bibfield
  {journal} {\bibinfo  {journal} {Phys. Rev. X}\ }\textbf {\bibinfo {volume}
  {8}},\ \bibinfo {pages} {041044} (\bibinfo {year} {2018})}\BibitemShut
  {NoStop}%
\bibitem [{\citenamefont {Kilaj}\ \emph {et~al.}(2018)\citenamefont {Kilaj},
  \citenamefont {Gao}, \citenamefont {R{\"o}sch}, \citenamefont {Rivero},
  \citenamefont {K{\"u}pper},\ and\ \citenamefont
  {Willitsch}}]{kilaj2018observation}%
  \BibitemOpen
  \bibfield  {author} {\bibinfo {author} {\bibfnamefont {A.}~\bibnamefont
  {Kilaj}}, \bibinfo {author} {\bibfnamefont {H.}~\bibnamefont {Gao}}, \bibinfo
  {author} {\bibfnamefont {D.}~\bibnamefont {R{\"o}sch}}, \bibinfo {author}
  {\bibfnamefont {U.}~\bibnamefont {Rivero}}, \bibinfo {author} {\bibfnamefont
  {J.}~\bibnamefont {K{\"u}pper}}, \ and\ \bibinfo {author} {\bibfnamefont
  {S.}~\bibnamefont {Willitsch}},\ }\href@noop {} {\bibfield  {journal}
  {\bibinfo  {journal} {Nat. Commun.}\ }\textbf {\bibinfo {volume} {9}}
  (\bibinfo {year} {2018})}\BibitemShut {NoStop}%
\bibitem [{\citenamefont {Puri}\ \emph {et~al.}(2019)\citenamefont {Puri},
  \citenamefont {Mills}, \citenamefont {Simbotin}, \citenamefont {Montgomery},
  \citenamefont {C{\^o}t{\'e}}, \citenamefont {Schneider}, \citenamefont
  {Suits},\ and\ \citenamefont {Hudson}}]{puri2019reaction}%
  \BibitemOpen
  \bibfield  {author} {\bibinfo {author} {\bibfnamefont {P.}~\bibnamefont
  {Puri}}, \bibinfo {author} {\bibfnamefont {M.}~\bibnamefont {Mills}},
  \bibinfo {author} {\bibfnamefont {I.}~\bibnamefont {Simbotin}}, \bibinfo
  {author} {\bibfnamefont {J.~A.}\ \bibnamefont {Montgomery}}, \bibinfo
  {author} {\bibfnamefont {R.}~\bibnamefont {C{\^o}t{\'e}}}, \bibinfo {author}
  {\bibfnamefont {C.}~\bibnamefont {Schneider}}, \bibinfo {author}
  {\bibfnamefont {A.~G.}\ \bibnamefont {Suits}}, \ and\ \bibinfo {author}
  {\bibfnamefont {E.~R.}\ \bibnamefont {Hudson}},\ }\href@noop {} {\bibfield
  {journal} {\bibinfo  {journal} {Nat. Chem.}\ }\textbf {\bibinfo {volume}
  {11}},\ \bibinfo {pages} {615} (\bibinfo {year} {2019})}\BibitemShut
  {NoStop}%
\bibitem [{\citenamefont {Hu}\ \emph {et~al.}(2021)\citenamefont {Hu},
  \citenamefont {Liu}, \citenamefont {Nichols}, \citenamefont {Zhu},
  \citenamefont {Qu{\'e}m{\'e}ner}, \citenamefont {Dulieu},\ and\ \citenamefont
  {Ni}}]{hu2020product}%
  \BibitemOpen
  \bibfield  {author} {\bibinfo {author} {\bibfnamefont {M.-G.}\ \bibnamefont
  {Hu}}, \bibinfo {author} {\bibfnamefont {Y.}~\bibnamefont {Liu}}, \bibinfo
  {author} {\bibfnamefont {M.~A.}\ \bibnamefont {Nichols}}, \bibinfo {author}
  {\bibfnamefont {L.}~\bibnamefont {Zhu}}, \bibinfo {author} {\bibfnamefont
  {G.}~\bibnamefont {Qu{\'e}m{\'e}ner}}, \bibinfo {author} {\bibfnamefont
  {O.}~\bibnamefont {Dulieu}}, \ and\ \bibinfo {author} {\bibfnamefont {K.-K.}\
  \bibnamefont {Ni}},\ }\href {\doibase 10.1038/s41557-020-00610-0} {\bibfield
  {journal} {\bibinfo  {journal} {Nat. Chem.}\ }\textbf {\bibinfo {volume}
  {13}},\ \bibinfo {pages} {435} (\bibinfo {year} {2021})}\BibitemShut
  {NoStop}%
\bibitem [{\citenamefont {Hu}\ \emph {et~al.}(2019)\citenamefont {Hu},
  \citenamefont {Liu}, \citenamefont {Grimes}, \citenamefont {Lin},
  \citenamefont {Gheorghe}, \citenamefont {Vexiau}, \citenamefont
  {Bouloufa-Maafa}, \citenamefont {Dulieu}, \citenamefont {Rosenband},\ and\
  \citenamefont {Ni}}]{hu2019direct}%
  \BibitemOpen
  \bibfield  {author} {\bibinfo {author} {\bibfnamefont {M.-G.}\ \bibnamefont
  {Hu}}, \bibinfo {author} {\bibfnamefont {Y.}~\bibnamefont {Liu}}, \bibinfo
  {author} {\bibfnamefont {D.}~\bibnamefont {Grimes}}, \bibinfo {author}
  {\bibfnamefont {Y.-W.}\ \bibnamefont {Lin}}, \bibinfo {author} {\bibfnamefont
  {A.}~\bibnamefont {Gheorghe}}, \bibinfo {author} {\bibfnamefont
  {R.}~\bibnamefont {Vexiau}}, \bibinfo {author} {\bibfnamefont
  {N.}~\bibnamefont {Bouloufa-Maafa}}, \bibinfo {author} {\bibfnamefont
  {O.}~\bibnamefont {Dulieu}}, \bibinfo {author} {\bibfnamefont
  {T.}~\bibnamefont {Rosenband}}, \ and\ \bibinfo {author} {\bibfnamefont
  {K.-K.}\ \bibnamefont {Ni}},\ }\href@noop {} {\bibfield  {journal} {\bibinfo
  {journal} {Science}\ }\textbf {\bibinfo {volume} {366}},\ \bibinfo {pages}
  {1111} (\bibinfo {year} {2019})}\BibitemShut {NoStop}%
\bibitem [{\citenamefont {Mayle}\ \emph {et~al.}(2012)\citenamefont {Mayle},
  \citenamefont {Ruzic},\ and\ \citenamefont {Bohn}}]{mayle2012statistical}%
  \BibitemOpen
  \bibfield  {author} {\bibinfo {author} {\bibfnamefont {M.}~\bibnamefont
  {Mayle}}, \bibinfo {author} {\bibfnamefont {B.~P.}\ \bibnamefont {Ruzic}}, \
  and\ \bibinfo {author} {\bibfnamefont {J.~L.}\ \bibnamefont {Bohn}},\
  }\href@noop {} {\bibfield  {journal} {\bibinfo  {journal} {Phys. Rev. A}\
  }\textbf {\bibinfo {volume} {85}},\ \bibinfo {pages} {062712} (\bibinfo
  {year} {2012})}\BibitemShut {NoStop}%
\bibitem [{\citenamefont {Mayle}\ \emph {et~al.}(2013)\citenamefont {Mayle},
  \citenamefont {Qu\'em\'ener}, \citenamefont {Ruzic},\ and\ \citenamefont
  {Bohn}}]{mayle2013scattering}%
  \BibitemOpen
  \bibfield  {author} {\bibinfo {author} {\bibfnamefont {M.}~\bibnamefont
  {Mayle}}, \bibinfo {author} {\bibfnamefont {G.}~\bibnamefont {Qu\'em\'ener}},
  \bibinfo {author} {\bibfnamefont {B.~P.}\ \bibnamefont {Ruzic}}, \ and\
  \bibinfo {author} {\bibfnamefont {J.~L.}\ \bibnamefont {Bohn}},\ }\href@noop
  {} {\bibfield  {journal} {\bibinfo  {journal} {Phys. Rev. A}\ }\textbf
  {\bibinfo {volume} {87}},\ \bibinfo {pages} {012709} (\bibinfo {year}
  {2013})}\BibitemShut {NoStop}%
\bibitem [{\citenamefont {Bonnet}\ and\ \citenamefont
  {C.~Rayez}(1999)}]{bonnet1999some}%
  \BibitemOpen
  \bibfield  {author} {\bibinfo {author} {\bibfnamefont {L.}~\bibnamefont
  {Bonnet}}\ and\ \bibinfo {author} {\bibfnamefont {J.}~\bibnamefont
  {C.~Rayez}},\ }\href@noop {} {\bibfield  {journal} {\bibinfo  {journal}
  {Phys. Chem. Chem. Phys.}\ }\textbf {\bibinfo {volume} {1}},\ \bibinfo
  {pages} {2383} (\bibinfo {year} {1999})}\BibitemShut {NoStop}%
\bibitem [{\citenamefont {Christianen}\ \emph
  {et~al.}(2019{\natexlab{a}})\citenamefont {Christianen}, \citenamefont
  {Zwierlein}, \citenamefont {Groenenboom},\ and\ \citenamefont
  {Karman}}]{christianen2019photoinduced}%
  \BibitemOpen
  \bibfield  {author} {\bibinfo {author} {\bibfnamefont {A.}~\bibnamefont
  {Christianen}}, \bibinfo {author} {\bibfnamefont {M.~W.}\ \bibnamefont
  {Zwierlein}}, \bibinfo {author} {\bibfnamefont {G.~C.}\ \bibnamefont
  {Groenenboom}}, \ and\ \bibinfo {author} {\bibfnamefont {T.}~\bibnamefont
  {Karman}},\ }\href@noop {} {\bibfield  {journal} {\bibinfo  {journal} {Phys.
  Rev. Lett.}\ }\textbf {\bibinfo {volume} {123}},\ \bibinfo {pages} {123402}
  (\bibinfo {year} {2019}{\natexlab{a}})}\BibitemShut {NoStop}%
\bibitem [{\citenamefont {Liu}\ \emph {et~al.}(2020{\natexlab{b}})\citenamefont
  {Liu}, \citenamefont {Hu}, \citenamefont {Nichols}, \citenamefont {Grimes},
  \citenamefont {Karman}, \citenamefont {Guo},\ and\ \citenamefont
  {Ni}}]{liu2020photo}%
  \BibitemOpen
  \bibfield  {author} {\bibinfo {author} {\bibfnamefont {Y.}~\bibnamefont
  {Liu}}, \bibinfo {author} {\bibfnamefont {M.-G.}\ \bibnamefont {Hu}},
  \bibinfo {author} {\bibfnamefont {M.~A.}\ \bibnamefont {Nichols}}, \bibinfo
  {author} {\bibfnamefont {D.~D.}\ \bibnamefont {Grimes}}, \bibinfo {author}
  {\bibfnamefont {T.}~\bibnamefont {Karman}}, \bibinfo {author} {\bibfnamefont
  {H.}~\bibnamefont {Guo}}, \ and\ \bibinfo {author} {\bibfnamefont {K.-K.}\
  \bibnamefont {Ni}},\ }\href@noop {} {\bibfield  {journal} {\bibinfo
  {journal} {Nat. Phys.}\ }\textbf {\bibinfo {volume} {16}},\ \bibinfo {pages}
  {1131} (\bibinfo {year} {2020}{\natexlab{b}})}\BibitemShut {NoStop}%
\bibitem [{\citenamefont {Gregory}\ \emph {et~al.}(2020)\citenamefont
  {Gregory}, \citenamefont {Blackmore}, \citenamefont {Bromley},\ and\
  \citenamefont {Cornish}}]{gregory2020loss}%
  \BibitemOpen
  \bibfield  {author} {\bibinfo {author} {\bibfnamefont {P.~D.}\ \bibnamefont
  {Gregory}}, \bibinfo {author} {\bibfnamefont {J.~A.}\ \bibnamefont
  {Blackmore}}, \bibinfo {author} {\bibfnamefont {S.~L.}\ \bibnamefont
  {Bromley}}, \ and\ \bibinfo {author} {\bibfnamefont {S.~L.}\ \bibnamefont
  {Cornish}},\ }\href@noop {} {\bibfield  {journal} {\bibinfo  {journal} {Phys.
  Rev. Lett.}\ }\textbf {\bibinfo {volume} {124}},\ \bibinfo {pages} {163402}
  (\bibinfo {year} {2020})}\BibitemShut {NoStop}%
\bibitem [{\citenamefont {Croft}\ and\ \citenamefont
  {Bohn}(2014)}]{croft2014long}%
  \BibitemOpen
  \bibfield  {author} {\bibinfo {author} {\bibfnamefont {J.~F.~E.}\
  \bibnamefont {Croft}}\ and\ \bibinfo {author} {\bibfnamefont {J.~L.}\
  \bibnamefont {Bohn}},\ }\href@noop {} {\bibfield  {journal} {\bibinfo
  {journal} {Phys. Rev. A}\ }\textbf {\bibinfo {volume} {89}},\ \bibinfo
  {pages} {012714} (\bibinfo {year} {2014})}\BibitemShut {NoStop}%
\bibitem [{\citenamefont {Li}\ \emph {et~al.}(2020)\citenamefont {Li},
  \citenamefont {Zhao}, \citenamefont {Xie},\ and\ \citenamefont
  {Guo}}]{li2020advances}%
  \BibitemOpen
  \bibfield  {author} {\bibinfo {author} {\bibfnamefont {J.}~\bibnamefont
  {Li}}, \bibinfo {author} {\bibfnamefont {B.}~\bibnamefont {Zhao}}, \bibinfo
  {author} {\bibfnamefont {D.}~\bibnamefont {Xie}}, \ and\ \bibinfo {author}
  {\bibfnamefont {H.}~\bibnamefont {Guo}},\ }\href@noop {} {\bibfield
  {journal} {\bibinfo  {journal} {The Journal of Physical Chemistry Letters}\
  }\textbf {\bibinfo {volume} {11}},\ \bibinfo {pages} {8844} (\bibinfo {year}
  {2020})}\BibitemShut {NoStop}%
\bibitem [{\citenamefont {Light}(1967)}]{light1967statistical}%
  \BibitemOpen
  \bibfield  {author} {\bibinfo {author} {\bibfnamefont {J.~C.}\ \bibnamefont
  {Light}},\ }\href@noop {} {\bibfield  {journal} {\bibinfo  {journal}
  {Discuss. Faraday Soc.}\ }\textbf {\bibinfo {volume} {44}},\ \bibinfo {pages}
  {14} (\bibinfo {year} {1967})}\BibitemShut {NoStop}%
\bibitem [{\citenamefont {Pechukas}(1976)}]{pechukas1976statistical}%
  \BibitemOpen
  \bibfield  {author} {\bibinfo {author} {\bibfnamefont {P.}~\bibnamefont
  {Pechukas}},\ }in\ \href@noop {} {\emph {\bibinfo {booktitle} {Dynamics of
  molecular collisions}}}\ (\bibinfo  {publisher} {Springer},\ \bibinfo {year}
  {1976})\ pp.\ \bibinfo {pages} {269--322}\BibitemShut {NoStop}%
\bibitem [{\citenamefont {Nikitin}\ and\ \citenamefont
  {Umanskii}(2012)}]{nikitin2012theory}%
  \BibitemOpen
  \bibfield  {author} {\bibinfo {author} {\bibfnamefont {E.~E.}\ \bibnamefont
  {Nikitin}}\ and\ \bibinfo {author} {\bibfnamefont {S.~Y.}\ \bibnamefont
  {Umanskii}},\ }\href@noop {} {\emph {\bibinfo {title} {Theory of slow atomic
  collisions}}},\ Vol.~\bibinfo {volume} {30}\ (\bibinfo  {publisher} {Springer
  Science \& Business Media},\ \bibinfo {year} {2012})\BibitemShut {NoStop}%
\bibitem [{\citenamefont {Liu}\ \emph {et~al.}(2021)\citenamefont {Liu},
  \citenamefont {Hu}, \citenamefont {Nichols}, \citenamefont {Yang},
  \citenamefont {Xie}, \citenamefont {Guo},\ and\ \citenamefont
  {Ni}}]{liu2020precision}%
  \BibitemOpen
  \bibfield  {author} {\bibinfo {author} {\bibfnamefont {Y.}~\bibnamefont
  {Liu}}, \bibinfo {author} {\bibfnamefont {M.-G.}\ \bibnamefont {Hu}},
  \bibinfo {author} {\bibfnamefont {M.~A.}\ \bibnamefont {Nichols}}, \bibinfo
  {author} {\bibfnamefont {D.}~\bibnamefont {Yang}}, \bibinfo {author}
  {\bibfnamefont {D.}~\bibnamefont {Xie}}, \bibinfo {author} {\bibfnamefont
  {H.}~\bibnamefont {Guo}}, \ and\ \bibinfo {author} {\bibfnamefont {K.-K.}\
  \bibnamefont {Ni}},\ }\href@noop {} {\bibfield  {journal} {\bibinfo
  {journal} {Nature}\ }\textbf {\bibinfo {volume} {593}},\ \bibinfo {pages}
  {379} (\bibinfo {year} {2021})}\BibitemShut {NoStop}%
\bibitem [{\citenamefont {Croft}\ \emph {et~al.}(2017)\citenamefont {Croft},
  \citenamefont {Makrides}, \citenamefont {Li}, \citenamefont {Petrov},
  \citenamefont {Kendrick}, \citenamefont {Balakrishnan},\ and\ \citenamefont
  {Kotochigova}}]{croft2017universality}%
  \BibitemOpen
  \bibfield  {author} {\bibinfo {author} {\bibfnamefont {J.}~\bibnamefont
  {Croft}}, \bibinfo {author} {\bibfnamefont {C.}~\bibnamefont {Makrides}},
  \bibinfo {author} {\bibfnamefont {M.}~\bibnamefont {Li}}, \bibinfo {author}
  {\bibfnamefont {A.}~\bibnamefont {Petrov}}, \bibinfo {author} {\bibfnamefont
  {B.}~\bibnamefont {Kendrick}}, \bibinfo {author} {\bibfnamefont
  {N.}~\bibnamefont {Balakrishnan}}, \ and\ \bibinfo {author} {\bibfnamefont
  {S.}~\bibnamefont {Kotochigova}},\ }\href@noop {} {\bibfield  {journal}
  {\bibinfo  {journal} {Nat. Commun.}\ }\textbf {\bibinfo {volume} {8}},\
  \bibinfo {pages} {15897} (\bibinfo {year} {2017})}\BibitemShut {NoStop}%
\bibitem [{\citenamefont {Kendrick}\ \emph {et~al.}(2021)\citenamefont
  {Kendrick}, \citenamefont {Li}, \citenamefont {Li}, \citenamefont
  {Kotochigova}, \citenamefont {Croft},\ and\ \citenamefont
  {Balakrishnan}}]{Kendrick2021}%
  \BibitemOpen
  \bibfield  {author} {\bibinfo {author} {\bibfnamefont {B.~K.}\ \bibnamefont
  {Kendrick}}, \bibinfo {author} {\bibfnamefont {H.}~\bibnamefont {Li}},
  \bibinfo {author} {\bibfnamefont {M.}~\bibnamefont {Li}}, \bibinfo {author}
  {\bibfnamefont {S.}~\bibnamefont {Kotochigova}}, \bibinfo {author}
  {\bibfnamefont {J.~F.~E.}\ \bibnamefont {Croft}}, \ and\ \bibinfo {author}
  {\bibfnamefont {N.}~\bibnamefont {Balakrishnan}},\ }\href@noop {} {\bibfield
  {journal} {\bibinfo  {journal} {Phys. Chem. Chem. Phys.}\ }\textbf {\bibinfo
  {volume} {23}},\ \bibinfo {pages} {5096} (\bibinfo {year}
  {2021})}\BibitemShut {NoStop}%
\bibitem [{\citenamefont {Karman}(2020)}]{PrivateCommunication}%
  \BibitemOpen
  \bibfield  {author} {\bibinfo {author} {\bibfnamefont {T.}~\bibnamefont
  {Karman}},\ }\href@noop {} {\bibfield  {journal} {\bibinfo  {journal}
  {Private communication}\ } (\bibinfo {year} {2020})}\BibitemShut {NoStop}%
\bibitem [{\citenamefont {Christianen}\ \emph
  {et~al.}(2019{\natexlab{b}})\citenamefont {Christianen}, \citenamefont
  {Karman},\ and\ \citenamefont {Groenenboom}}]{ChristianenDOS2019}%
  \BibitemOpen
  \bibfield  {author} {\bibinfo {author} {\bibfnamefont {A.}~\bibnamefont
  {Christianen}}, \bibinfo {author} {\bibfnamefont {T.}~\bibnamefont {Karman}},
  \ and\ \bibinfo {author} {\bibfnamefont {G.~C.}\ \bibnamefont
  {Groenenboom}},\ }\href@noop {} {\bibfield  {journal} {\bibinfo  {journal}
  {Phys. Rev. A}\ }\textbf {\bibinfo {volume} {100}},\ \bibinfo {pages}
  {032708} (\bibinfo {year} {2019}{\natexlab{b}})}\BibitemShut {NoStop}%
\bibitem [{\citenamefont {Levine}(2009)}]{levine2009molecular}%
  \BibitemOpen
  \bibfield  {author} {\bibinfo {author} {\bibfnamefont {R.~D.}\ \bibnamefont
  {Levine}},\ }\href@noop {} {\emph {\bibinfo {title} {Molecular Reaction
  Dynamics}}}\ (\bibinfo  {publisher} {Cambridge University Press},\ \bibinfo
  {year} {2009})\BibitemShut {NoStop}%
\bibitem [{\citenamefont {Ni}\ \emph {et~al.}(2008)\citenamefont {Ni},
  \citenamefont {Ospelkaus}, \citenamefont {De~Miranda}, \citenamefont {Pe'Er},
  \citenamefont {Neyenhuis}, \citenamefont {Zirbel}, \citenamefont
  {Kotochigova}, \citenamefont {Julienne}, \citenamefont {Jin},\ and\
  \citenamefont {Ye}}]{ni2008high}%
  \BibitemOpen
  \bibfield  {author} {\bibinfo {author} {\bibfnamefont {K.-K.}\ \bibnamefont
  {Ni}}, \bibinfo {author} {\bibfnamefont {S.}~\bibnamefont {Ospelkaus}},
  \bibinfo {author} {\bibfnamefont {M.}~\bibnamefont {De~Miranda}}, \bibinfo
  {author} {\bibfnamefont {A.}~\bibnamefont {Pe'Er}}, \bibinfo {author}
  {\bibfnamefont {B.}~\bibnamefont {Neyenhuis}}, \bibinfo {author}
  {\bibfnamefont {J.}~\bibnamefont {Zirbel}}, \bibinfo {author} {\bibfnamefont
  {S.}~\bibnamefont {Kotochigova}}, \bibinfo {author} {\bibfnamefont
  {P.}~\bibnamefont {Julienne}}, \bibinfo {author} {\bibfnamefont
  {D.}~\bibnamefont {Jin}}, \ and\ \bibinfo {author} {\bibfnamefont
  {J.}~\bibnamefont {Ye}},\ }\href@noop {} {\bibfield  {journal} {\bibinfo
  {journal} {Science}\ }\textbf {\bibinfo {volume} {322}},\ \bibinfo {pages}
  {231} (\bibinfo {year} {2008})}\BibitemShut {NoStop}%
\bibitem [{\citenamefont {Seto}\ \emph {et~al.}(2000)\citenamefont {Seto},
  \citenamefont {Le~Roy}, \citenamefont {Verges},\ and\ \citenamefont
  {Amiot}}]{seto2000direct}%
  \BibitemOpen
  \bibfield  {author} {\bibinfo {author} {\bibfnamefont {J.~Y.}\ \bibnamefont
  {Seto}}, \bibinfo {author} {\bibfnamefont {R.~J.}\ \bibnamefont {Le~Roy}},
  \bibinfo {author} {\bibfnamefont {J.}~\bibnamefont {Verges}}, \ and\ \bibinfo
  {author} {\bibfnamefont {C.}~\bibnamefont {Amiot}},\ }\href@noop {}
  {\bibfield  {journal} {\bibinfo  {journal} {J. Chem. Phys.}\ }\textbf
  {\bibinfo {volume} {113}},\ \bibinfo {pages} {3067} (\bibinfo {year}
  {2000})}\BibitemShut {NoStop}%
\bibitem [{\citenamefont {Ospelkaus}\ \emph
  {et~al.}(2010{\natexlab{b}})\citenamefont {Ospelkaus}, \citenamefont {Ni},
  \citenamefont {Qu\'em\'ener}, \citenamefont {Neyenhuis}, \citenamefont
  {Wang}, \citenamefont {de~Miranda}, \citenamefont {Bohn}, \citenamefont
  {Ye},\ and\ \citenamefont {Jin}}]{OspelkausHyperfine2010}%
  \BibitemOpen
  \bibfield  {author} {\bibinfo {author} {\bibfnamefont {S.}~\bibnamefont
  {Ospelkaus}}, \bibinfo {author} {\bibfnamefont {K.-K.}\ \bibnamefont {Ni}},
  \bibinfo {author} {\bibfnamefont {G.}~\bibnamefont {Qu\'em\'ener}}, \bibinfo
  {author} {\bibfnamefont {B.}~\bibnamefont {Neyenhuis}}, \bibinfo {author}
  {\bibfnamefont {D.}~\bibnamefont {Wang}}, \bibinfo {author} {\bibfnamefont
  {M.~H.~G.}\ \bibnamefont {de~Miranda}}, \bibinfo {author} {\bibfnamefont
  {J.~L.}\ \bibnamefont {Bohn}}, \bibinfo {author} {\bibfnamefont
  {J.}~\bibnamefont {Ye}}, \ and\ \bibinfo {author} {\bibfnamefont {D.~S.}\
  \bibnamefont {Jin}},\ }\href@noop {} {\bibfield  {journal} {\bibinfo
  {journal} {Phys. Rev. Lett.}\ }\textbf {\bibinfo {volume} {104}},\ \bibinfo
  {pages} {030402} (\bibinfo {year} {2010}{\natexlab{b}})}\BibitemShut
  {NoStop}%
\bibitem [{\citenamefont {Aldegunde}\ \emph {et~al.}(2008)\citenamefont
  {Aldegunde}, \citenamefont {Rivington}, \citenamefont {\ifmmode~\dot{Z}\else
  \.{Z}\fi{}uchowski},\ and\ \citenamefont {Hutson}}]{Aldegunde2008}%
  \BibitemOpen
  \bibfield  {author} {\bibinfo {author} {\bibfnamefont {J.}~\bibnamefont
  {Aldegunde}}, \bibinfo {author} {\bibfnamefont {B.~A.}\ \bibnamefont
  {Rivington}}, \bibinfo {author} {\bibfnamefont {P.~S.}\ \bibnamefont
  {\ifmmode~\dot{Z}\else \.{Z}\fi{}uchowski}}, \ and\ \bibinfo {author}
  {\bibfnamefont {J.~M.}\ \bibnamefont {Hutson}},\ }\href {\doibase
  10.1103/PhysRevA.78.033434} {\bibfield  {journal} {\bibinfo  {journal} {Phys.
  Rev. A}\ }\textbf {\bibinfo {volume} {78}},\ \bibinfo {pages} {033434}
  (\bibinfo {year} {2008})}\BibitemShut {NoStop}%
\bibitem [{Sup()}]{Supplement}%
  \BibitemOpen
  \href@noop {} {\bibinfo  {journal} {Supplemental material}\ }\BibitemShut
  {NoStop}%
\bibitem [{\citenamefont {Li}\ \emph {et~al.}(2019)\citenamefont {Li},
  \citenamefont {Li}, \citenamefont {Makrides}, \citenamefont {Petrov},\ and\
  \citenamefont {Kotochigova}}]{LiUniversalLoss2019}%
  \BibitemOpen
\bibfield  {journal} {  }\bibfield  {author} {\bibinfo {author} {\bibfnamefont
  {H.}~\bibnamefont {Li}}, \bibinfo {author} {\bibfnamefont {M.}~\bibnamefont
  {Li}}, \bibinfo {author} {\bibfnamefont {C.}~\bibnamefont {Makrides}},
  \bibinfo {author} {\bibfnamefont {A.}~\bibnamefont {Petrov}}, \ and\ \bibinfo
  {author} {\bibfnamefont {S.}~\bibnamefont {Kotochigova}},\ }\href@noop {}
  {\bibfield  {journal} {\bibinfo  {journal} {Atoms}\ }\textbf {\bibinfo
  {volume} {7}},\ \bibinfo {pages} {36} (\bibinfo {year} {2019})}\BibitemShut
  {NoStop}%
\bibitem [{\citenamefont {Gersema}\ \emph {et~al.}(2021)\citenamefont
  {Gersema}, \citenamefont {Voges}, \citenamefont {zum Alten~Borgloh},
  \citenamefont {Koch}, \citenamefont {Hartmann}, \citenamefont {Zenesini},
  \citenamefont {Ospelkaus}, \citenamefont {Lin}, \citenamefont {He},\ and\
  \citenamefont {Wang}}]{gersema2021probing}%
  \BibitemOpen
  \bibfield  {author} {\bibinfo {author} {\bibfnamefont {P.}~\bibnamefont
  {Gersema}}, \bibinfo {author} {\bibfnamefont {K.~K.}\ \bibnamefont {Voges}},
  \bibinfo {author} {\bibfnamefont {M.~M.}\ \bibnamefont {zum Alten~Borgloh}},
  \bibinfo {author} {\bibfnamefont {L.}~\bibnamefont {Koch}}, \bibinfo {author}
  {\bibfnamefont {T.}~\bibnamefont {Hartmann}}, \bibinfo {author}
  {\bibfnamefont {A.}~\bibnamefont {Zenesini}}, \bibinfo {author}
  {\bibfnamefont {S.}~\bibnamefont {Ospelkaus}}, \bibinfo {author}
  {\bibfnamefont {J.}~\bibnamefont {Lin}}, \bibinfo {author} {\bibfnamefont
  {J.}~\bibnamefont {He}}, \ and\ \bibinfo {author} {\bibfnamefont
  {D.}~\bibnamefont {Wang}},\ }\href@noop {} {\bibfield  {journal} {\bibinfo
  {journal} {arXiv:2103.00510}\ } (\bibinfo {year} {2021})}\BibitemShut
  {NoStop}%
\bibitem [{\citenamefont {Bause}\ \emph {et~al.}(2021)\citenamefont {Bause},
  \citenamefont {Schindewolf}, \citenamefont {Tao}, \citenamefont {Duda},
  \citenamefont {Chen}, \citenamefont {Quéméner}, \citenamefont {Karman},
  \citenamefont {Christianen}, \citenamefont {Bloch},\ and\ \citenamefont
  {Luo}}]{bause2021collisions}%
  \BibitemOpen
  \bibfield  {author} {\bibinfo {author} {\bibfnamefont {R.}~\bibnamefont
  {Bause}}, \bibinfo {author} {\bibfnamefont {A.}~\bibnamefont {Schindewolf}},
  \bibinfo {author} {\bibfnamefont {R.}~\bibnamefont {Tao}}, \bibinfo {author}
  {\bibfnamefont {M.}~\bibnamefont {Duda}}, \bibinfo {author} {\bibfnamefont
  {X.-Y.}\ \bibnamefont {Chen}}, \bibinfo {author} {\bibfnamefont
  {G.}~\bibnamefont {Quéméner}}, \bibinfo {author} {\bibfnamefont
  {T.}~\bibnamefont {Karman}}, \bibinfo {author} {\bibfnamefont
  {A.}~\bibnamefont {Christianen}}, \bibinfo {author} {\bibfnamefont
  {I.}~\bibnamefont {Bloch}}, \ and\ \bibinfo {author} {\bibfnamefont {X.-Y.}\
  \bibnamefont {Luo}},\ }\href@noop {} {\bibfield  {journal} {\bibinfo
  {journal} {arXiv:2103.00889}\ } (\bibinfo {year} {2021})}\BibitemShut
  {NoStop}%
\bibitem [{\citenamefont {Baranov}\ \emph {et~al.}(2012)\citenamefont
  {Baranov}, \citenamefont {Dalmonte}, \citenamefont {Pupillo},\ and\
  \citenamefont {Zoller}}]{baranov2012condensed}%
  \BibitemOpen
  \bibfield  {author} {\bibinfo {author} {\bibfnamefont {M.~A.}\ \bibnamefont
  {Baranov}}, \bibinfo {author} {\bibfnamefont {M.}~\bibnamefont {Dalmonte}},
  \bibinfo {author} {\bibfnamefont {G.}~\bibnamefont {Pupillo}}, \ and\
  \bibinfo {author} {\bibfnamefont {P.}~\bibnamefont {Zoller}},\ }\href@noop {}
  {\bibfield  {journal} {\bibinfo  {journal} {Chem. Rev.}\ }\textbf {\bibinfo
  {volume} {112}},\ \bibinfo {pages} {5012} (\bibinfo {year}
  {2012})}\BibitemShut {NoStop}%
\bibitem [{\citenamefont {DeMille}(2002)}]{demille2002quantum}%
  \BibitemOpen
  \bibfield  {author} {\bibinfo {author} {\bibfnamefont {D.}~\bibnamefont
  {DeMille}},\ }\href@noop {} {\bibfield  {journal} {\bibinfo  {journal} {Phys.
  Rev. Lett.}\ }\textbf {\bibinfo {volume} {88}},\ \bibinfo {pages} {067901}
  (\bibinfo {year} {2002})}\BibitemShut {NoStop}%
\bibitem [{\citenamefont {Ni}\ \emph {et~al.}(2018)\citenamefont {Ni},
  \citenamefont {Rosenband},\ and\ \citenamefont {Grimes}}]{ni2018dipolar}%
  \BibitemOpen
  \bibfield  {author} {\bibinfo {author} {\bibfnamefont {K.-K.}\ \bibnamefont
  {Ni}}, \bibinfo {author} {\bibfnamefont {T.}~\bibnamefont {Rosenband}}, \
  and\ \bibinfo {author} {\bibfnamefont {D.~D.}\ \bibnamefont {Grimes}},\
  }\href@noop {} {\bibfield  {journal} {\bibinfo  {journal} {Chem. Sci.}\
  }\textbf {\bibinfo {volume} {9}},\ \bibinfo {pages} {6830} (\bibinfo {year}
  {2018})}\BibitemShut {NoStop}%
\bibitem [{\citenamefont {Safronova}\ \emph {et~al.}(2018)\citenamefont
  {Safronova}, \citenamefont {Budker}, \citenamefont {DeMille}, \citenamefont
  {Kimball}, \citenamefont {Derevianko},\ and\ \citenamefont
  {Clark}}]{Safronova2018Review}%
  \BibitemOpen
  \bibfield  {author} {\bibinfo {author} {\bibfnamefont {M.~S.}\ \bibnamefont
  {Safronova}}, \bibinfo {author} {\bibfnamefont {D.}~\bibnamefont {Budker}},
  \bibinfo {author} {\bibfnamefont {D.}~\bibnamefont {DeMille}}, \bibinfo
  {author} {\bibfnamefont {D.~F.~J.}\ \bibnamefont {Kimball}}, \bibinfo
  {author} {\bibfnamefont {A.}~\bibnamefont {Derevianko}}, \ and\ \bibinfo
  {author} {\bibfnamefont {C.~W.}\ \bibnamefont {Clark}},\ }\href@noop {}
  {\bibfield  {journal} {\bibinfo  {journal} {Rev. Mod. Phys.}\ }\textbf
  {\bibinfo {volume} {90}},\ \bibinfo {pages} {025008} (\bibinfo {year}
  {2018})}\BibitemShut {NoStop}%
\end{thebibliography}

\clearpage
\setcounter{equation}{0}
\renewcommand{\theequation}{S\arabic{equation}}
\setcounter{figure}{0}
\renewcommand{\thefigure}{S\arabic{figure}}
\setcounter{table}{0}
\renewcommand{\thetable}{S\arabic{table}}
\onecolumngrid
\begin{center}
\large{\textbf{Supplementary Materials: \\ Detection of Long-Lived Complexes in Ultracold Atom-Molecule Collisions}}
\end{center}
\twocolumngrid

\section{Experimental Setup}
Detailed descriptions of the sample preparation and ion detection scheme can be found in previous works~\cite{hu2019direct,liu2020probing,liu2020photo}. In short, the KRb molecules populate a single hyperfine state, $\kt{m_{I}^{\text{K}}=-4,m_{I}^{\text{Rb}}=1/2}$, of the absolute ground electronic, vibrational, and rotational state of the molecules, $\kt{X^{1}\Sigma^{+},v=0,N=0}$, and the Rb atoms populate the lowest energy atomic hyperfine state, $\kt{F=1,m_{F}=1}$. Here, $m_{I}, v,$ and $N$ represent the projection of the nuclear spin onto the quantization axis set by an external magnetic field, the molecular vibrational quantum number, and the molecular rotational quantum number, respectively, and $F$ and $m_{F}$ represent the total atomic angular momentum and its projection onto the quantization axis, respectively. The initial sample preparation is performed in a $1064$ nm, continuously operated crossed ODT of total intensity 11.3 kW/cm$^2$, at a magnetic field of $B=542$ G. Control over the Rb atom number density is achieved by selectively removing different amounts of the excess Rb atoms which remain in the trap after the molecular gas has been created. The maximum atomic density utilized in the experiment, $1.6\times10^{12}$ cm$^{-3}$, is obtained, for example, by leaving these atoms untouched, and the pure molecular sample is obtained by removing all of these atoms.

To count the molecule number remaining in the system as a function of time for the data shown in Fig. 2 of the main text, we perform absorption imaging on the molecules. Specifically, we first hold the atom-molecule mixture in the crossed ODT at $B=542$ G for the desired amount of time. We then remove the excess Rb atoms from the system by transferring them to the $\kt{F=2,m_{F}=2}$ state with frequency-swept microwaves, and applying a pulse of resonant light. After all the Rb atoms have been removed, we selectively dissociate those ground state molecules remaining in the original hyperfine state, $\kt{m_{I}^{\text{K}}=-4,m_{I}^{\text{Rb}}=1/2}$, and perform absorption imaging on the resulting free atoms to extract the total molecule number.

For the ion data shown in Figs. 3 and 4 of the main text, after the initial atom-molecule mixture has been created at $B=542$ G, we lower the magnetic field over $30$ ms to a value of $B=30$ G, and turn on an electric field $E=17$ V/cm (Fig. 1 inset). The electric field is required to extract ions from the sample region, and the magnetic field, which is needed to maintain a quantization axis, must be lowered to allow for ion detection. Once the fields are turned on, the ODT intensity modulation is applied, along with the UV ionization laser pulses, until the sample is depleted (${\sim}1$ s).

In Fig.~3(a) of the main text, we show the dependence of the measured KRb$_{2}^{+}$ ion signal on the initial Rb density, $n_{a}$. There, we observed a saturation of this signal with increasing values of $n_{a}$. This effect is the result of a competition between the finite UV pulse repetition frequency and the decay rate of the mixture. If the decay rate arising from atom-molecule collisions is too high, for example, fewer ions can be collected, given the finite sampling frequency, before the sample is depleted.

\section{Photo-ionization Scheme}
To photo-ionize the KRb$_{2}^{*}$ collision complexes, we utilize a pulsed UV laser which has a $7$ ns pulse duration and operates at $354.85$ nm. Because of the negligible lab-frame translational energy and transient nature of the KRb$_{2}^{*}$ complexes, they reside within the same region of the ODT as the trapped sample. We therefore shape the UV ionization laser into a Gaussian beam that overlaps with the atomic and molecular clouds, and which has $1/e^2$ waist diameters of $280$ $\mu$m and $175$ $\mu$m along the two orthogonal beam axes. Since the UV photon energy at $354.85$ nm is below the ionization thresholds of both KRb and Rb, the overlap of the ionization laser with the trapped sample does not result in a measurable depletion of either the atoms or the molecules in the mixture. In addition, as we detect no KRb$_{2}^{+}$ counts in the pure molecular sample (Fig.~3(a)), it follows that the KRb$_{2}^{+}$ signal observed in the main text is not the result of dissociative ionization of intermediate K$_{2}$Rb$_{2}^{*}$ complexes formed through reactive collisions of KRb molecules. This effect was previously observed for UV wavelengths $\lesssim345$ nm in Ref.~\cite{hu2019direct}.

The photon-ionization threshold for the ground rovibronic state of the neutral KRb$_{2}$ complex, which sits $1608$ cm$^{-1}$ below the incident energy of the atom-molecule collision channel (Fig.~1), coincides with a photon energy of $377.6$ nm~\cite{hu2019direct}. The corresponding photon-ionization threshold for the transient KRb$_{2}^{*}$ complexes at the incident collision energy is ${\sim}402$ nm. Because of this, the $354.85$ nm ionization wavelength utilized here has a sufficient photon energy to ionize both transient and deeply bound states of the complex. To confirm that the KRb$_{2}^{+}$ ion signal observed in the main text arises from transient collision complexes (KRb$_{2}^{*}$), and not from more deeply bound complexes, we have also performed measurements analogous to those shown in Fig.~4(a) of the main text using a pulsed UV ionization laser operating at $394.81$ nm. At this ionization wavelength, the KRb$_{2}^{+}$ ion signal demonstrates the same behavior as that shown in Fig.~4(a), and we observe no measurable difference between the corresponding KRb$_{2}^{*}$ complex lifetimes. However, the overall strength of the KRb$_{2}^{+}$ ion signal at this wavelength is relatively reduced due to limitations in the available $394.81$ nm laser power. We therefore find it sufficient for the purposes of this work to operate the pulsed ionization laser at $354.85$ nm, which provides a higher time-averaged laser power.

\section{ODT Intensity Modulation Scheme}
For the data shown in Figs. 3 and 4 of the main text, we apply a $1.5$ kHz square-wave intensity modulation with a $25\%$ duty cycle to the H and V ODTs using acousto-optic modulators (AOMs). Both beams are always modulated synchronously, so that they follow the same timing. We do this in order to probe the KRb$_{2}^{*}$ population at lower, controllable $1064$ nm intensity levels without noticeably altering the thermodynamic properties of the atom-molecule mixture. That is, because the trapping frequencies, $f_{\text{trap}}$, of both the molecular and atomic gases satisfy $f_{\text{trap}}<0.4$ kHz along all three principal axes of the trap, the chosen modulation frequency for the time-averaged potential does not notably affect the temperature or density of the atoms and molecules in the sample.

To further ensure that the temperature and density of the atomic and molecular gases in the intensity-modulated trap configuration are the same as those in the continuous-wave (CW) ODT, we make sure that the time-averaged intensities of both the H and V beams are equal to their continuously operated levels. For the H ODT, which is always modulated at full depth, the peak intensity during the ``high" phase of its modulation period is $22.8$ kW/cm$^2$. This is four times its CW level, $5.7$ kW/cm$^2$, so that the $25\%$ modulation duty cycle is taken into account. For the V ODT, the modulation depth can be varied in order to control the $1064$ nm intensity level which is present when we apply the UV ionization laser pulse. To keep the time-averaged intensity of the V ODT fixed at its CW level, $5.6$ kW/cm$^2$, we utilize the constraint, $(3I + I')/4=5.6$ kW/cm$^2$, where $I$ is the intensity level of the ``low" phase of the modulation, and $I'$ is that of the ``high" phase of the modulation (Fig.~3 inset).

By varying the value of $I$, we can change the level of the total $1064$ nm intensity at the point within the modulation period where the UV ionization pulse is applied. This total intensity can be expressed as $I_{tot}=I+I_{H,leak}$, where $I_{H,leak}=5.01$ W/cm$^{2}$ represents the leakage intensity from the H ODT during the dark phase of its modulation. This is a constant for each value of $I$ examined, and is the result of imperfect suppression of the H ODT light by the AOM modulation scheme used in the experiment.

\section{Intensity Calibration}
In the experimental setup, the H and V ODTs are formed from $1064$ nm Gaussian laser beams with $1/e^2$ waist diameters of $70$ and $200$ $\mu$m, respectively, which intersect one another at an angle of approximately $70^{\circ}$ (Fig.~1 inset). Both beams are derived from the same $1064$ nm laser source, which has a spectral width of $1$ kHz. In the continuously operated state of the crossed ODT, the typical total $1064$ nm intensity is $11.3$ kW/cm$^2$. The cigar-shaped atomic and molecular clouds resulting from this trap configuration have $2\sigma$ Gaussian widths of $8$, $8$, and $38$ $\mu$m, and $6$, $6$, and $28$ $\mu$m, respectively, along the three principal axes of the trap. Therefore, the variation of the $1064$ nm intensity over these widths is less than $7\%$, and is considered to be constant across the sample. As such, the total optical intensities utilized throughout the main text are calibrated using the peak intensities of the Gaussian beams.

\section{Inelastic Atom-Molecule Collisions}
If the total angular momentum, $\vec{J}=\vec{L}+\vec{N}$, is conserved in collisions between KRb and Rb, then inelastic collisions which flip the KRb nuclear spins should be suppressed given the particular combination of atomic and molecular hyperfine states used in the experiment. Here, $L$ is the orbital angular momentum of the collision. Because the molecules and atoms are distinguishable particles, they are restricted to collide via $s$-wave scattering at ultralow temperatures, so that there is no initial orbital angular momentum ($L=0$). As the molecules are prepared in their rotational ground state ($N=0$), there is also no rotational angular momentum. The initial total angular momentum is therefore $J=0$.

To understand which final combinations of atomic and molecular internal states might be accessible through inelastic atom-molecule collisions, one must determine which combinations have a total energy that is less than or equal to that of the incoming channel. Because the KRb molecules are initially prepared in $N=0$, and the energy splitting between the $N=0$ and $N=1$ manifolds is $>2$ GHz~\cite{Aldegunde2008,OspelkausHyperfine2010}, the final rotational state of the molecules must also be $N=0$. Additionally, because Rb is prepared in the lowest energy hyperfine state, $\kt{F=1,m_{F}=1}$, and the energy splitting between neighboring $m_{F}$ states at the magnetic fields used in the experiment ($>20$ MHz at $B=30$ G) is much greater than the energy spread of the molecular hyperfine states within the $N=0$ rotational manifold ($<200$ kHz at $B=30$ G), the Rb atoms must remain in $\kt{F=1,m_{F}=1}$ after the collisions. Therefore, the only energetically accessible collision channels are those which leave the Rb hyperfine state unchanged, but which flip the KRb nuclear spins. However, the only molecular hyperfine states that are lower in energy relative to the initial state, $\kt{m_{I}^{\text{K}}=-4,m_{I}^{\text{Rb}}=1/2}$, are those that have a larger summed nuclear spin projection $m_{F}^{\text{KRb}} \equiv m_{I}^{\text{K}} + m_{I}^{\text{Rb}}$ (e.g. $\kt{m_{I}^{\text{K}}=-4,m_{I}^{\text{Rb}}=3/2}$)~\cite{Aldegunde2008,OspelkausHyperfine2010}. Thus, a finite angular momentum must be imparted to the molecule to increase the value of $m_{F}^{\text{KRb}}$ and lower the internal energy. As both $N$ and the atomic hyperfine state remain unchanged due to energetic constraints, this momentum must come from $\vec{L}$ of the post-collision molecules and atoms. If $J=0$ is conserved, however, which is equivalent in this case to the conservation of $L$ since $N=0$ remains unchanged, then the final value of $L$ must also be $L=0$. In this case, there is no angular momentum available to flip the KRb nuclear spins and lower the internal energy, so that the molecules must remain in their original hyperfine state after a collision. In this sense, total (or orbital) angular momentum conservation implies that the atom-molecule mixture used in the experiment should be stable against inelastic, spin-changing collisions.

\section{Field dependence of the complex lifetime}
The presence of external electric ($E$) or magnetic ($B$) fields in the experiment could potentially lead to a breakdown in the conservation of total angular momentum, $\vec{J}=\vec{L}+\vec{N}$, throughout the KRb-Rb collisions. This, in turn, could increase the DOS of the KRb$_{2}^{*}$ collision complex, and therefore also the KRb$_{2}^{*}$ lifetime, by several orders of magnitude~\cite{ChristianenDOS2019}. For this reason, we have experimentally examined the effect of these fields on the complex lifetime by performing measurements analogous to those shown in Fig.~4 of the main text using different values of $E$ and $B$. The results are summarized in Table~\ref{table1}, and the relative orientations of the externally applied fields are shown in Fig.~1 of the main text.

To ensure that our observations are sensitive to field-dependent changes in the complex lifetime, we measure the KRb$_{2}^{*}$ population dynamics at these different $E-$ and $B-$fields in the presence of a finite intensity of ODT light, $31.7$ W/cm$^2$, which corresponds to the red curve in Fig.~4(a) of the main text. Due to the experimental constraints of our ion detection scheme, we cannot lower the electric field below $17$ V/cm without losing the KRb$_{2}^{+}$ ion signal. For the same reasons, we cannot measure the complex lifetime at higher magnetic fields (e.g. $B=300$ G) without simultaneously increasing the value of the electric field. However, within the range of fields accessible to our experiment, we observe no significant variation in the characteristic growth rate, $R=\tau_{c}^{-1}+\beta_{1}I_{tot}$, of the KRb$_{2}^{*}$ population at a fixed value of the total $1064$ nm intensity, $I_{tot}=31.7$ W/cm$^2$ (Table~\ref{table1}). This indicates that the KRb$_{2}^{*}$ complex lifetime is not significantly affected by an order of magnitude increase in the external field strengths. Further investigations are required, however, to determine if the complex lifetime changes in any way in the total absence of these external fields.

\begin{table}[ht]
\caption{\label{table1}
Measured growth rate, $R$, of the KRb$_{2}^{*}$ population after a rapid change in the total $1064$ nm intensity to a value $I_{tot}=31.7$ W/cm$^2$. The measurements are performed in the presence of different electric ($E$) and magnetic ($B$) field strengths, but otherwise use the same experimental procedures that were utilized for the data shown in Fig. 4 of the main text.
}
\begin{ruledtabular}
\begin{tabular}{l c c c}
\textrm{$E$ (V/cm)}&
\textrm{$B$ (G)}&
\textrm{$I_{tot}$ (W/cm$^2$)}&
\textrm{Growth rate $R$ ($\mu$s$^{-1}$)}\\
\colrule
17 & 30 & 31.7 & 0.018(2)\\
343 & 30 & 31.7 & 0.018(2)\\
343 & 300 & 31.7 & 0.017(2)\\
\end{tabular}
\end{ruledtabular}
\end{table}

\end{document}